\documentclass[onecolumn]{emulateapj}
\lefthead{PEPPER \& GAUDI} \righthead{CLUSTER TRANSIT SEARCHES}
\def\eq#1{equation (\ref{#1})}

\def\ntr{N_{tr}}
\def\ttr{t_{tr}}
\def\teq{t_{eq}}
\def\tn{t_{night}}
\def\pt{{\cal P}_{t}}
\def\pw{{\cal P}_{W}}

\def\psn{{\cal P}_{S/N}}
\def\ceq{\chi^2_{eq}}
\def\cmin{\Delta\chi^2_{min}}
\def\nn{N_{n}}

\def\ptot{{\cal P}_{tot}}
\def\fdet{f_{det}}
\def\sn{{\rm S/N}}
\def\ct{{\cal T}}

\begin{document}

\title{Searching for Transiting Planets in Stellar Systems}
\author 
{Joshua Pepper\altaffilmark{1} and B.\ Scott Gaudi\altaffilmark{2} }
\altaffiltext{1}{Ohio State University Department of Astronomy, 4055 McPherson Lab, 140 West 18th Ave., Columbus, OH 43210}
\altaffiltext{2}{Harvard-Smithsonian Center for Astrophysics, 60 Garden St., Cambridge, MA 02138}
\email{pepper@astronomy.ohio-state.edu, sgaudi@cfa.harvard.edu}

\begin{abstract}
We analyze the properties of searches devoted to finding planetary
transits by observing simple stellar systems, such as globular
clusters, open clusters, and the Galactic bulge.  We develop the
analytic tools necessary to predict the number of planets that a
survey will detect as a function of the parameters of the system (age,
extinction, distance, richness, mass function), the observational
setup (nights observed, bandpass, exposure time, telescope diameter,
detector characteristics), site properties (seeing, sky background), 
and the planet properties (frequency, period, and radius).

We find that for typical parameters, the detection probability is
maximized for $I$-band observations.  At fixed planet period and
radius, the signal-to-noise ratio of a planetary transit in the $I$-band is
weakly dependent on the mass of the primary for sources with flux
above the sky background, and falls very sharply for sources below
sky.  Therefore, for typical targets, the number of detectable planets
is roughly proportional to the number of stars with transiting planets
with fluxes above sky (and not necessarily the number of sources with
photometric error less a given threshold).  Furthermore, for rising
mass functions, the majority of the planets will be detected around
sources with fluxes near sky.  In order to maximize
the number of detections, experiments should therefore be
tailored such that sources near sky are above the required detection
threshold.  Once this requirement is met, the number of detected
planets is relatively weakly dependent on the detection threshold,
diameter of the telescope, exposure time, seeing, age of the
system, and planet radius, for typical ranges of these parameters
encountered in current transit searches in stellar systems.  The
number of detected planets is a strongly decreasing function of the
distance to the system, implying that the nearest, richest clusters
may prove to be optimal targets.
\end{abstract}
\keywords{techniques: photometric -- surveys -- planetary systems}

\bigskip

\section{Introduction}

Although radial velocity (RV) searches have provided an enormous
amount of information about the ensemble properties of extrasolar
planets, the interpretation of these results has been somewhat
complicated by the fact that the planets' properties have been shaped
by the poorly-understood process of planetary migration.  Short-period
planets (periods $P\la 10~{\rm days}$, i.e.\ ``Hot Jupiters'') are
essential for understanding this phenomenon, since they have all
almost certainly reached their current positions via migration, and
because they are the easiest to detect via several methods, including
both radial velocities and transits.  Thus, it is possible to rapidly
acquire the statistics necessary for uncovering diagnostic trends in
their ensemble properties, which may provide clues to the physical
mechanisms that drive migration. Although RV searches have
been and will continue to be very successful in detecting these
planets, transit searches are rapidly gaining in importance.

There are currently over a dozen collaborations searching for planets
via transits (see \citealt{horne03}).  These searches have recently
started to come to fruition, and six close-in extrasolar giant planets
have been detected using the transit technique to date
\citep{konacki03a,bouchy04,pont04,konacki04,konacki05,alonso04}, with
many more likely to follow.  Notably, transit searches have already
uncovered a previously unknown population of ``Very Hot Jupiters'' --
massive planets with $P \la 3~{\rm days}$. Current transit searches
can be roughly divided into two categories.  Shallow surveys observe
bright ($V\la 14$) nearby stars with small aperture, large
field-of-view dedicated instruments
\citep{bakos04,kane04,boru01,pep04,alonso04,mc04,deeg04}.  The goal of
these surveys is primarily to find transiting planets around bright
stars, which facilitate the extensive follow-up studies that are
possible for transiting planets
\citep{charbonneau02,vidal03,vidal04,charbonneau05,deming05}.  On the
other hand, deep surveys monitor faint $(V\ga 14)$ stars using larger
aperture telescopes with small field-of-view instruments.  Typically,
these searches do not use dedicated facilities, and thus are generally
limited to campaigns lasting for a few weeks.  In contrast to the
shallow surveys, deep surveys will find planets around stars that are
too faint for all but the most rudimentary reconnaissance.  However,
the primary advantage of these searches is that a large number of
stars can be simultaneously probed for transiting planets. This allows
such surveys to detect relatively rare planets, as well as probe
planets in very different environments, and so robustly constrain the
statistics of close-in planets.  Deep searches can be further
subdivided into two categories, namely searches around field stars in
the Galactic plane
\citep{udalski02a,udalski02b,udalski02c,udalski03,udalski04,mo03}, and
searches toward simple stellar systems.

Simple stellar systems, such as globular clusters, open clusters, and
the Galactic bulge, are excellent laboratories for transit surveys, as
they provide a relatively uniform sample of $\sim 10^{3-5}$ stars of
the same age, metallicity, and distance.  Furthermore, such surveys
have several important advantages over field surveys.  With minimal
auxiliary observations, stellar systems
provide independent estimates for the mass and radius of the target
stars through main-sequence fitting to color-magnitude diagrams.  An
independent estimate for the stellar mass and radius, even with a
crude transit light curve, can allow one to completely characterize
the system parameters (assuming a circular orbit and a negligible
companion mass).  Transit data alone, without knowledge of the
properties of the host stars, does not allow for breaking of the
degeneracy between the stellar and planet radius and orbital
semi-major axis.  As a result, considerable additional expenditure of
resources is required to confirm the planetary nature of transit
candidates from field surveys
\citep{dreizler02,konacki03b,pont05a,bouchy05,gallardo05}.
Furthermore, using the results of field transit surveys to place
constraints on the ensemble properties of close-in planets is hampered
by a lack of information about the properties of the population of
host stars, as well as strong biases in the observed distributions of
planetary parameters relative to the underlying intrinsic planet
population \citep{gsm05,pont05a,gaudi05,dorsher05}.  In contrast, the
biases encountered in surveys toward stellar systems are considerably
less severe, and furthermore are easily quantified because the
properties of the host stars are known.  This allows for accurate
calibration of the detection efficiency of a particular survey, and so
enables robust inferences about the population of planets from the
detection (or lack thereof) of individual planetary companions
\citep{gilliland00,weldrake05,mo05,burke05}.

There are a number of projects devoted to searching for transiting planets in
stellar systems
\citep{gilliland00,burke03,street03,bruntt03,dk04,vb05,mo05,weldrake05,hidas05}.
These projects have observed or are observing a number of different kinds of systems,
with various ages, metallicities, and distances, using a variety of
observing parameters, such as telescope aperture and observing
cadence.  Although several authors have discussed general
considerations in designing and executing optimal surveys toward
stellar systems \citep{janes96,vb05,gaudi00}, these studies have been
somewhat fractured, and primarily qualitative in nature.  To date
there has been no rigorous, quantitative, and comprehensive
determination of how the different characteristics of the target
system and observing parameters affect the number of transiting
planets one would expect to find.  To this end, here we develop an
analytic model of transit surveys toward simple, homogeneous stellar
systems.  This model is useful for understanding the basic properties
of such surveys, for predicting the yield of a particular survey, as
well as for establishing guidelines that observers can use to make
optimum choices when observing particular targets.

We concentrate on the simplest model that incorporates the majority of
the important features of transit surveys toward stellar systems.  We
consider simple systems containing main-sequence stars of the same age
and metallicity.  We ignore the effects of weather, systematic errors
(except at the most rudimentary level), and variations in seeing and 
background.  Although we feel our analysis captures the basic
properties of such searches without considering these effects, it is
straightforward to extend our model to include these and other
real-world effects.

In \S \ref{sec:genform} we develop the equations and overall formalism
that we use to characterize the detection probabilities of certain
planets in specific systems with a given observational setup.  In \S
\ref{sec:analapprox} we describe various analytic approximations
we use to make sense of our detailed calculations, and we show how the
transit detection probabilities depend on stellar mass and the
characteristics of a particular survey.  In \S \ref{sec:AddIngr} we
list various physical relations and numerical approximations we use to
calculate detection probabilities.  In \S \ref{sec:results} we describe
the dependence of the detection probabilities on the input parameters, 
and we present an application of our results in
\S \ref{sec:app}.  We summarize and conclude in \S\ref{sec:con}.

\bigskip

\section{General Formalism} \label{sec:genform}

\bigskip

\subsection{The Number of Detected Transiting Planets}

For a given stellar system, the number of transiting planets with
periods between $P$ and $P+dP$ and radii between $r$ and $r+dr$ that
can be detected around stars with masses between $M$ and $M+dM$ is,
\begin{equation} \label{eqn:dN}
\frac{d^3N_{det}}{dMdrdP} = N_* f_p \frac{d^2p}{drdP} \ptot(M,P,r)
\frac{dn}{dM}.
\end{equation}
Here, $N_{det}$ is the number of detected transiting planets; $N_*$ is
the total number of stars in the system; $d^2p/drdP$ is the
probability that a planet around a star in the system has a period
between $P$ and $P+dP$ and a radius between $r$ and $r+dr$; $f_p$ is
the fraction of stars in the system with planets; $\ptot(M,P,r)$ is
the probability that a planet of radius $r$ and orbital period $P$
will be detected around a star of mass $M$; and ${ d}n/{ d}M$ is
the mass function of the stars in the system.

There are a number of assumptions that enter into \eq{eqn:dN}:
\begin{itemize}
\item We assume that $f_p$ and $d^2p/drdP$ are independent of
$M$.  We normalize $d^2p/drdP$ to unity over a specific range of
planetary radii and periods, and normalize $dn/dM$ to unity over a
specific range of stellar masses.  Therefore, $N_*$ is the number of
(single) stars in the mass range of interest, and $f_p$ is the fraction of such
stars harboring planets in the range of planetary radii and periods of
interest.  The number of such planets is thus $N_p=f_p N_*$, and the
fraction that are detected is $\fdet\equiv N_{det}/N_p$.  The
normalization of ${ d}n/{ d}M$ is described in \S
\ref{sec:massfunc}, and the normalization of $d^2p/drdP$ is described
in \S \ref{sec:PlanDist}.
\item We choose to use $P$ as our independent parameter rather than
semi-major axis $a$, since it is the more directly observable quantity
in transit searches, and it simplifies the following discussion
considerably.
\item We note that one of the primary simplifying assumptions in
\eq{eqn:dN} is that all the target stars are at the same distance from
the observer, which is an excellent
assumption for most stellar systems. 
\end{itemize}

\bigskip

\subsection{Detection Probabilities $\pt$, $\pw$, $\psn$ \label{sec:dps}}

Following \citet{gaudi00}, we separate $\ptot(M,P,r)$ into 
three factors
\begin{equation}
\ptot(M,P,r) = \pt(M,P) \psn(M,P,r) \pw(P),
\label{eqn:pall}
\end{equation}
where $\pt$ is the probability that a planet transits its parent star,
$\psn$ is the probability that, should a transit occur during a night
of observing, it will yield a Signal-to-Noise ratio (S/N) that is
higher than some threshold value, and $\pw$ is the window function
which describes the probability that more than one transit will occur
during the observations.

\bigskip

\subsubsection{Transit Probability $\pt$ \label{sec:pt}}

The probability that a planet will transit its parent star
is simply
\begin{equation}
\pt = \frac{R}{a} = 
\left(\frac{4 \pi^2}{G}\right)^{1/3} M^{-1/3} R P^{-2/3}.
\label{eqn:pt}
\end{equation}
This form of $\pt$ assumes that the planet is in a circular orbit.  
We will make this assumption throughout this paper. 

\bigskip

\subsubsection{Window Probability $\pw$\label{sec:pw}}

\begin{figure}[t]
\epsscale{1.0}
\plotone{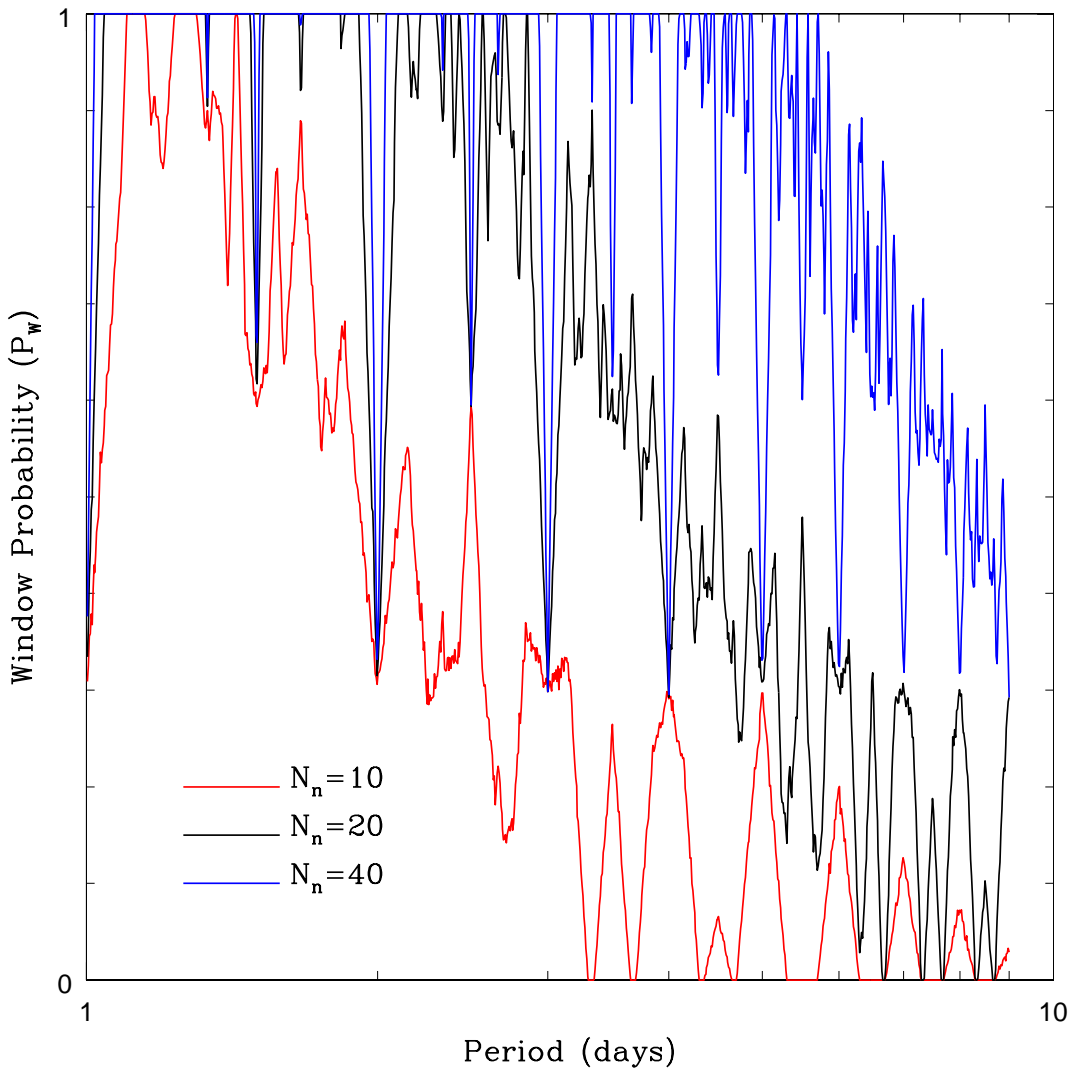}
\caption{The window probability $\pw$, which is the probability
that a planet with a given period will exhibit $n$ separate
transits during the times observations are made.  
Here we have assumed 7.2 observable hours each night, that the run 
is $\nn$ nights long, and we have required $n=2$ transits to occur during the observing window.
The lines show the results for
$\nn=10$ (dotted), 20 (solid), and 40 (dashed).}
\label{fig:win}
\end{figure}

The window function $\pw(P)$ quantifies the probability that a planet
with a given period $P$ will exhibit $n$ different transits during the
times when observations are made.  See \citet{gaudi00} for a
mathematical definition of $\pw$.  We will consider observational
campaigns from single sites comprising a total of $\nn$ contiguous
nights of length $\tn$.  For an exploration of the effects
of alternate observing strategies on $\pw$, we refer the reader to 
a comprehensive discussion by \citet{vb05}. We will assume that no time is lost to
weather.  Finally, we require only that the center of the transit
occurs during the night; therefore $\pw$ depends only on $n$, $\nn$,
$\tn$, and $P$, and does not depend on the transit duration.  Note
that our definition differs slightly from the definition by
\citet{mo03}.  In Figure
(\ref{fig:win}), we show $\pw$ as a function of $P$ for $\nn=10, 20, 40$ nights and
$\tn=7.2$ hours, for the requirement of $n=2$ transits (which we will require
throughout).

\bigskip

\subsubsection{Signal-to-Noise Probability $\psn$\label{sec:psn}}

In this section, we determine $\psn$, the probability that a single
transit will exceed a $\sn$ value 
larger than some minimum threshold $\sn$ value.\footnote{By folding an observed light curve
about the proper period, it is possible to improve the total $\sn$ 
over that of a single transit by $\sim n^{1/2}$, where $n$
is the number of transits occurring when observations are made.  We have chosen
a more conservative approach of requiring a minimum $\sn$ based on a single
transit because, for observational campaigns such as those typically
considered here, the probability of seeing many transits is low, and furthermore
detailed and well-sampled individual transit signals are crucial for distinguishing
bona fide transits from false positives.  In Appendix \ref{app:totsn}, we rederive
the results of this section for the alternative detection criterion
based on the total $\sn$ of folded transit light curves.  The difference
between these two approaches is relatively minor for the surveys
considered here, although the total $\sn$ approach favors short period
planets more heavily.}  The signal-to-noise of a single transit
is $\sn=(\Delta \chi^2)^{1/2}$, where $\Delta\chi^2$ is the difference
in $\chi^2$ between a constant flux and a transit fit to the data.
For simplicity, we will model all transits as boxcar curves.  
In this case, and under the assumption that only a small fraction of the data points 
occur during transit, the $\Delta \chi^2$ of a transit is simply,
\begin{equation}
\Delta\chi^2_{ tr}=N_{tr} \left( \frac{\delta}{\sigma} \right)^{2}.
\label{eqn:chi}
\end{equation}
Here $N_{tr}$ is the number of observations during the transit,
$\delta$ is the fractional change in the star's brightness during the
transit, and $\sigma$ is the fractional error of an individual flux
measurement.

The number of observations $N_{tr}$ during a transit is related to
the observing timescales: $N_{tr} = t_{tr}/( t_{read} +
t_{exp} )$, where $t_{tr}$ is the duration of the transit,
$t_{read}$ is read time of the detector, and $t_{exp}$ is
exposure time.\footnote{This model assumes that all transits are 
observed from beginning to end.  We consider the effects of
partial transits in Appendix
\ref{app:PartTrans}.}  We can put $t_{tr}$ in terms of the
fundamental parameters: 
\begin{equation}  \label{eqn:ttran}
t_{tr} = 2x \sqrt{ \frac{a}{GM} } = x \left( \frac{4 P}{\pi GM} \right)^{1/3}.
\end{equation}
Here, $x$ is half the length of the 
chord that traces the path of the transiting planet across the face of the 
star.  Geometrically, $x = R\sqrt{1 - b^2}$ where $b$ is the 
impact parameter of the transit.  That is, $b R$ is equal to the 
distance from the equator to the latitude of the transit.  For a transit 
with an inclination of $90$ degrees, $x = R$ and $b = 0$, while for a 
grazing eclipse $x$ is nearly 0 and $b = 1$.  
We define $t_{eq}$ as the duration of an equatorial
transit (i.e. $t_{eq} = t_{tr}(b = 0)$), and therefore
 $t_{tr} = t_{eq}\sqrt{1 - b^2}$.

We assume a transit will be discovered if and only if $\Delta\chi^2_{tr}$ 
is larger than some threshold value $\Delta\chi^2_{min}$.  We note
that $\Delta\chi^2_{tr} = \Delta\chi^2_{eq}\sqrt{1 - b^2}$.
Therefore, the probability of achieving sufficient S/N is essentially
a step function, such that:
\begin{equation}  \label{eqn:psn1a}
\frac{d\psn}{db} = \Theta \left[ \Delta\chi_{eq}^2\sqrt{1
- b^2} - \Delta\chi_{min}^2 \right],
\end{equation}
where $\Theta$ is the step function ($\Theta(x) = 0$ for $x < 0$;
$\Theta(x) = 1$ for $x \ge 0$).  Equation (\ref{eqn:psn1a}) provides
us with the probability that a transit with impact parameter between
$b$ and $b+db$ will yield a sufficient S/N to be
detected.  This can be determined for a given set of intrinsic
parameters of the system ($M, r,$ and $P$) and the observational
parameters which we will list later.

We can assume that the impact parameters of transiting systems are distributed 
uniformly.  We will take $b$ as our fundamental test of S/N, so that if 
a transit in a system with a given set of intrinsic parameters achieves 
sufficient S/N to be detected with an equatorial transit $b = 0$, then 
it will be also detectable with any $b$ up to some inclination 
$b_{max}$, beyond which point it will not achieve sufficient S/N.  We will 
integrate equation (\ref{eqn:psn1a}) from $b = 0$ to $b_{max}$, which is 
the range over which $\psn = 1$:
\begin{equation}  \label{eqn:psn1b}
\psn= \int_0^1 \frac{d\psn}{db} db = \int_0^1 \Theta \left[ 
\Delta\chi_{eq}^2\sqrt{1 - b^2} -  \Delta\chi_{min}^2 \right] db \, .
\end{equation}
This formulation makes it easy to eliminate 
the step function, since any argument with $b > b_{max}$ will cause 
the argument of the step function to be less than 0, and so the value of the 
integrand will equal 0.  We can therefore integrate $\psn$ from 0 
to $b_{max}$, so the left hand side of equation (\ref{eqn:psn1b}) becomes
\begin{equation}  \label{eqn:psn1c}
\psn = \int_0^{b_{max}} db = b_{max} \, .
\end{equation}
We can take the right hand side of equation (\ref{eqn:psn1b}) and note that the argument of 
the step function will equal 0 when evaluated at $b_{max}$.  Setting 
$\Delta\chi_{eq}^2\sqrt{1 - b_{max}^2} -  \Delta\chi_{min}^2 = 0$ and solving for $b_{max}$, we 
then have
\begin{equation}  \label{eqn:beta1a}
\psn = b_{max} = \sqrt{1- \left( \frac{\Delta\chi^2_{min}}{\Delta\chi^2_{eq}} \right)^2}\, ,\qquad {\rm if}\, \Delta\chi^2_{min} \le \Delta \chi^2_{eq},
\end{equation}
and $\psn=0$ otherwise.

We must now determine the dependence of the various factors in equation 
(\ref{eqn:beta1a}) on the independent parameters $M, r,$ and $P$, as well as 
the observing parameters.  Using \eq{eqn:chi}, we can 
put $\Delta\chi_{eq}^2$ in terms of the independent parameters and 
$(\delta/\sigma)^2$.  We must therefore relate $(\delta/\sigma)^2$ to the 
independent parameters.  Assuming Poisson statistics,
$\sigma = \sqrt{N_{S} + N_{B}}/N_S$, 
where $N_{S}$ is the number of photons recorded from a target star in a 
given exposure, and $N_{B}$ is the number of background photons.  In terms 
of the observing parameters, $N_{S} = f_{\lambda} t_{exp} \pi (D/2)^2$, where 
$f_{\lambda}$ is the flux of photons with wavelength $\lambda$ from the target 
star, $D$ is the telescope aperture, and we have assumed a filled
aperture.  Flux is related to luminosity by
\begin{equation}  \label{eqn:fnu}
f_{\lambda} = \frac{L_{\lambda}}{4 \pi d^2} 10^{-A_{\lambda}/2.5} , 
\end{equation}
where $L_{\lambda}$ is the star's photon luminosity at wavelength $\lambda$, 
$A_{\lambda}$ is the interstellar extinction at wavelength $\lambda$, and 
$d$ is the distance to the system.  Turning to the background sky photons, we can define 
\begin{equation}  \label{eqn:NB}
N_{B} = S_{sky,\lambda} \Omega t_{exp} \pi (D/2)^2,
\end{equation}
where $S_{sky,\lambda}$ is the photon surface brightness of the sky in wavelength $\lambda$ 
and $\Omega$ is effective area of the seeing disk.

Putting all this together, we can write $\Delta\chi^2_{eq}$ in terms 
of the parameters of the planet, primary, and observational setup,
\begin{equation} \label{eqn:beta1b}
\Delta\chi^2_{eq} = 
(1024\pi)^{-1/3} 
\frac{t_{exp}}{t_{read}+t_{exp}}
\left(\frac{r}{R}\right)^4 
\left(\frac{D}{d}\right)^2 
\left(\frac{PR^3}{GM}\right)^{1/3} 
L_\lambda 10^{-0.4A_\lambda}
\left(1+\frac{S_{sky,\lambda}\Omega 4\pi d^2}{L_\lambda 10^{-0.4A_\lambda}}\right)^{-1}.
\end{equation}
This form can then be inserted into \eq{eqn:beta1a} to find $\psn$. 

Note that we have assumed Poisson statistics, no losses due to the
atmosphere, telescope, or instrumentation, and no additional
background flux other than that due to sky (no blending).  In \S\ref{sec:error},
we introduce a systematic floor to the photometric error $\sigma$.
However, other than this one concession to reality, our results
will represent the results of ideal, photon-limited experiments,
and are therefore in some sense the best case outcomes.  When designing actual
experiments, such real-world complications need to be considered carefully to 
ensure that they do not substantially alter the conclusions drawn here.  

\bigskip

\section{Analytic Approximations - Sensitivity as a Function of Primary Mass } \label{sec:analapprox}

To lowest order, the main-sequence
population of a coeval, homogeneous stellar system forms a 
one-parameter system of stars.  Therefore,
a novel aspect of transit searches in stellar systems is that, once the 
cluster, planet, and observational
parameters have been specified, the sensitivity
of different stars can be characterized by a single parameter,
namely the stellar mass.  This simple behavior, combined
with assumptions about the mass-luminosity relation, mass-radius
relation, and mass function, allows us derive analytic results for the 
sensitivity of transit surveys as a function of stellar mass.

Here we consider the sensitivity of a given transit search to planets
of a given radius $r$ and period $P$ as a function of the primary mass
$M$.  Adopting power-law forms for the mass-luminosity and mass-radius
relations,  we rewrite the analytic detection probabilities
for $\pt$ and $\psn$ that we derived in \S\ref{sec:dps} in terms of
$M$.  We note that, due to the manner in which we have defined it,
$\pw$ depends only on $P$ and the observational parameters, and not
on $M$.  This simplifies the understanding of the sensitivity
considerably, since $\pw$ is the only factor that must be calculated
numerically.  

\bigskip

\subsection{Mass-Luminosity and Mass-Radius Relations} \label{sec:mlmr}
 
We adopt generic power-law mass-luminosity and mass-radius
relations,
\begin{equation}\label{eqn:mrl}
R=R_\odot \left(\frac{M}{M_\odot}\right)^{\alpha}, \qquad L_\lambda =
L_{\lambda,\odot} \left(\frac{M}{M_\odot}\right)^{\beta_\lambda}
\end{equation}
where $L_{\lambda,\odot}$ is the photon luminosity at a wavelength
$\lambda$ for a solar-mass star.  The power-law index for the
mass-luminosity relation is wavelength-dependent, such
that the $\beta_\lambda$ index accounts for bolometric corrections
for particular bandpasses.

We note that neither empirically calibrated nor theoretically
predicted mass-radius and mass-luminosity relations are strict power
laws.  However, the power-law relations lead to useful analytic
results that aid in the intuitive results of the more precise results
presented later.  Furthermore, for stars near $M\sim M_\sun$ and
optical bandpasses, this approximation is reasonably accurate.

For the most part, we will keep the resulting analytic expressions in
terms of the variables $\alpha$ and $\beta_\lambda$, rather than
substitute specific values.  However, as will become clear, some
interesting properties of these expressions are seen for realistic
values of these parameters.  Therefore, where appropriate, we
occasionally insert numerical values for $\alpha$ and $\beta_\lambda$.
As we show later, for most targets, the $I$-band proves to be
optimal in terms of maximizing the signal-to-noise ratio of detected
transits.  For the $I$-band, and $0.3~M_\odot \la M \la 2 M_\odot$,
typical values are $\alpha =1$ and $\beta_I=3.5$.

\bigskip

\subsection{Dependence of $\psn$ on $M$\label{sec:psn_on_M}}

We first consider $\Delta\chi^2_{eq}$ and $\psn$.  
Substituting \eq{eqn:mrl} into equations (\ref{eqn:beta1a})
and (\ref{eqn:beta1b}), we find after some algebra,
\begin{equation}\label{eqn:chi2mass}
\frac{\Delta\chi^2_{eq}}{\Delta\chi^2_{min}}=\frac{1}{C_1}
\left(\frac{M}{M_\odot}\right)^{-(3\alpha-\beta_\lambda+1/3)}
\left[1+C_2 \left(\frac{M}{M_\odot}\right)^{-\beta_\lambda}\right]^{-1}
\end{equation}
\begin{equation}  \label{eqn:generic_psn}
\psn = \left\{1 - C_1^2 \left(\frac{M}{M_\odot}\right)^{2(3\alpha-\beta_\lambda+1/3)} 
\left[1+C_2 \left(\frac{M}{M_\odot}\right)^{-\beta_\lambda}\right]^2\right\}^{1/2}
\end{equation}
where we absorb all the constants and parameters except for mass into the 
new constants $C_1$ and $C_2$, which are given by,
\begin{equation}\label{eqn:c1}
C_1 = 
(1024\pi)^{1/3}\cmin
\left(1+\frac{t_{read}}{t_{exp}}\right)
\left(\frac{r}{R_\odot}\right)^{-4} 
\left(\frac{d}{D}\right)^{2} 
\left(\frac{GM_\odot}{P R_\odot^3L_{\lambda,\odot}^3}\right)^{1/3}
10^{0.4A_\lambda},
\end{equation}
\begin{equation}\label{eqn:c2}
C_2 =
\frac{4\pi d^2 S_{sky,\lambda}\Omega}{L_{\lambda,\odot}10^{-0.4A_\lambda}}.
\end{equation}
Note that $C_2$ is simply the ratio of the flux in the seeing disk to the flux
of a star of $M_\odot$. 

Inspection of the behavior of equation (\ref{eqn:generic_psn}) as a function
of $M$ reveals that there are two different regimes.  In the first
regime the second term within the square brackets is much smaller than unity
and hence negligible.  This
is the regime in which the photon noise is dominated by the source (i.e. the target star).   In the opposite
regime, where that term is much larger than unity, the noise is dominated by
the sky background.  The transition between these two regimes occurs at
the mass $M_{sky}$ where
the flux from the star is equal to the flux from the sky background,
\begin{equation}  \label{eqn:Mnoise}
M_{sky} = C_2^{1/\beta_\lambda} M_\odot.
\end{equation}

The behavior of $\psn$ as a function of mass depends on the value of
$\Delta\chi^2_{eq}/\Delta\chi^2_{min}$ at $M=M_{sky}$.  If
$\Delta\chi^2_{eq}/\Delta\chi^2_{min}<1$ at $M=M_{sky}$,
then the ability to detect planets is limited by the source noise for all
the stars in the system.  Conversely, if
$\Delta\chi^2_{eq}/\Delta\chi^2_{min}>1$ at $M=M_{sky}$,
then the ability to detect planets around the faintest
stars in the system is limited by noise due to the sky background. 
That is to say, a particular experiment can be characterized by
whether the flux of the faintest star around which a planet can be detected is
brighter or dimmer than the sky.  We call these the  
``source limited'' and ``background limited'' regimes, respectively.  
We shall see the implications of this distinction shortly.  An experiment is in the background
limited regime when
$\Delta\chi^2_{eq}/\Delta\chi^2_{min}\ge 1$
for $M\ge M_{sky}$, which implies,
\begin{equation}\label{eqn:regime}
2C_1C_2^{(3\alpha-\beta_\lambda+1/3)/\beta_\lambda}\le 1.
\end{equation}
In the source noise limited regime, we find that
\begin{equation}
\frac{\Delta\chi^2_{eq}}{\Delta\chi^2_{min}}=\frac{1}{C_1}
\left(\frac{M}{M_\odot}\right)^{-(3\alpha-\beta_\lambda+1/3)},\qquad{\rm (Source~Limited)},
\end{equation}
where $\psn$ becomes
\begin{equation}  \label{eqn:source_psn}
\psn \simeq  \left[1 - C_1^2 \left(\frac{M}{M_\odot}\right)^{2(3\alpha-\beta_\lambda+1/3)} \right]^{1/2},\qquad{\rm (Source~Limited)}.
\end{equation}
On the other hand, in the background noise limited regime, 
\begin{equation}\label{eqn:chi2mass_sky}
\frac{\Delta\chi^2_{eq}}{\Delta\chi^2_{min}}=\frac{1}{C_1C_2}
\left(\frac{M}{M_\odot}\right)^{-(3\alpha-2\beta_\lambda+1/3)},\qquad{\rm (Background~Limited)},
\end{equation}
and 
\begin{equation}  \label{eqn:sky_psn}
\psn \simeq \left[1 - (C_1 C_2)^2 \left(\frac{M}{M_\odot}\right)^{2(3\alpha - 2\beta_\lambda + 1/3)} \right]^{1/2},
\qquad{\rm(Background~Limited)} .
\end{equation}
Both of these equations have the same general form.  For masses below
a certain threshold, $M_{th}$, there is no chance of detecting a
transit.  The formula for $M_{th}$ can be determined separately
for the two different noise regimes.  In the source noise limited
regime, we have
\begin{equation}  \label{eqn:Mbreak_source}
M_{th,s} =
\left(C_1\right)^{-1/(3\alpha-\beta_\lambda+1/3)}M_\odot,
\end{equation}
while in the background noise limited regime,
\begin{equation}  \label{eqn:Mbreak_sky}
M_{th,b} = \left(C_1
C_2\right)^{-1/(3\alpha-2\beta_\lambda+1/3)}M_\odot.
\end{equation}

Thus $\psn$ as a function of stellar mass is approximately
a step function, and the placement of the step, $M_ {th}$ 
will depend on whether the faintest star around which a planet
is detectable (for which $\Delta\chi^2_{eq}>\Delta\chi^2_{min}$) is brighter or dimmer than the sky.  
Although the labels ``source limited'' and ``background 
limited'' refer to the faintest star for which a planet is detectable, 
and not to all the stars in the system, 
we shall see shortly that the integrated detection probability will depend 
primarily on the lowest-mass stars. 

It is highly instructive to insert numerical values for $\alpha$ and
$\beta_\lambda$ and consider the behavior of $\Delta\chi^2_{eq}$ and
$\psn$ in the source and background limited regimes.  Adopting values
appropriate to the $I$-band, ($\alpha=1$ and $\beta_I=3.5$), we have
$3\alpha-\beta_\lambda+1/3 = -1/6$, and thus $\Delta\chi^2_{eq}
\propto M^{1/6}$ in the source-noise limited
regime.  Thus, for sources above sky, the signal-to-noise
is an extremely weak function of mass.  On the other hand, for sources
below sky, we have that $3\alpha-2\beta_\lambda+1/3 = -11/3$, and thus
$\Delta\chi^2_{eq} \propto M^{11/3}$, an extremely strong function of
mass.  Taken together, these results imply that, if it is possible to
detect transiting planets around any stars in the target system, it is
possible to detect planets with the same radius and period around all
stars in the system above sky.  For stars fainter than sky, the
detection rapidly becomes impossible with decreasing mass.  These effects are illustrated in 
\S \ref{sec:hsres}.

These results have an interesting corollary that informs the 
experimental design.  If the experiment is background limited
(i.e.\ $\Delta\chi^2_{eq}(M_{sky}) \ge \Delta\chi^2_{min}$), then 
the minimum stellar mass around which a planet
is detectable is $M_{th,b}/M_\odot = (C_1C_2)^{3/11}$, whereas in the source 
limited regime $M_{th,s}/M_\odot = (C_1)^{6}$.  Since the constants $C_1$ and $C_2$
depend on the parameters of the target system, the experimental setup, and
the observational parameters, these scaling relations generally imply that the
yield of experiments
in the background limited regime is relatively insensitive to the precise 
values of these parameters, whereas the opposite
is true for experiments in the source limited regime.  Said very crudely:
specific experiments are either capable of detecting planets or they are not.
Experiments should be tailored such 
that $\Delta\chi^2_{eq}/\Delta\chi^2_{min}\ge 1$ at $M=M_{sky}$, which 
implies that $2C_1C_2^{(3\alpha-\beta_\lambda+1/3)/\beta_\lambda}\le 1$,
but provided 
this requirement is well-satisfied, changing the observational parameters will 
have little effect on the number of detected planets. 

\bigskip

\subsection{Dependence of $\pt$ on $M$}

We next consider $\pt$.  Substituting \eq{eqn:mrl} into \eq{eqn:beta1a}
and (\ref{eqn:beta1b}), 
\begin{equation}\label{eqn:ptmass}
\pt = C_3
\left(\frac{M}{M_\odot}\right)^{\alpha-1/3},
\end{equation}
where we have defined
\begin{equation}\label{eqn:c3}
C_3=\left( \frac{4\pi^2R_\odot^3}{P^2 GM_\odot}\right)^{1/3}=0.238\left(\frac{P}{\rm day}\right)^{-2/3}.
\end{equation}

\bigskip

\subsection{Dependence of $dn/dM$ on $M$}

We assume a differential mass function of the form
\begin{equation}\label{eqn:massfunction}
\frac{dn}{dM} \equiv k \left(\frac{M}{M_\odot}\right)^{\gamma}.
\end{equation}
The constant $k$ must be chosen such that that the integral over $dn/dM$ is equal to unity, i.e.\ such that
\begin{equation} \label{eqn:intmass}
\int^{M_{max}}_{M_{min}} dM k \left( \frac{M}{M_{\odot}} \right)^\gamma = 1 \, ,
\end{equation}
where $M_{max}$ and $M_{min}$ are the
masses of the largest and smallest stars in the system to be considered.
Solving \eq{eqn:intmass} 
for $k$ gives us $k = (\gamma + 1) M_\odot^\gamma / [ M_{max}^{\gamma + 1} - M_{min}^{\gamma + 1} ]$.

\bigskip

\subsection{Dependence of $\ptot(dn/dM)$ on $M$\label{sec:ptot_on_M}}

We can now use these forms for $\psn$ and $\pt$, together with
assumptions about the mass function of the stellar system $dn/dM$, to evaluate the detection sensitivity to planets with a given
set of properties.

To a first approximation, $\psn$ is simply a step function such that
$\psn = \Theta(M-M_{th})$, where $M_{th}$ is the minimum
threshold mass.  This is given by $M_{th}=M_{th,b}$ if 
$\Delta\chi^2(M_{sky}) \ge \Delta\chi^2_{min}$, and $M_{th}=M_{th,b}$ otherwise.
Thus for masses $M\ge M_{th}$, the sensitivity as a function of mass is
dominated by the effects of $\pt$ and $dn/dM$.  We can write
\begin{equation}\label{eqn:pnm}
\ptot(M,P,r) \frac{dn}{dM}= \pw(P) C_3(P) k \left(\frac{M}{M_\odot}\right)^{\zeta} \Theta(M-M_{th}).
\end{equation}
where $\zeta \equiv \alpha-1/3+\gamma$.  For a Saltpeter slope of 
$\gamma=-2.35$, and $\alpha=1$, $\zeta\simeq -1.68$. 
Therefore, under the assumption that the frequency of planets of 
a given radius and period is independent of the mass of the primary, the 
number of detected planets is dominated by parent stars with mass
near $M_{th}$, which, in the usual case of a background-dominated 
experiment, is for stars with flux just below the sky. 

\bigskip

\section{Additional Ingredients} \label{sec:AddIngr}

In \S\ref{sec:analapprox} we adopted several
simplifying assumptions and approximations that allowed us to derive
analytic expressions for the detectability of planets as a function of
primary mass.  Inspection of these expressions
 allowed us to infer some generic properties of
transit searches in stellar systems.  However, in order to make
realistic estimates of the number of planets a particular
survey will detect, here we add a few additional ingredients to the
basic formalism presented in \S\ref{sec:genform}.  We will also present a
somewhat more sophisticated treatment of the mass-luminosity relation,
as well as adopt specific values for several parameters as necessary to make
quantitative predictions.

\bigskip

\subsection{Reconsidering the Mass-Luminosity Relation} \label{sec:betabol}

The above analysis approximated the mass-luminosity 
relation as a simple power law in each wavelength band. 
As we have already discussed, this assumption is incorrect in 
detail.  We therefore provide a somewhat better approximation
to the mass-luminosity relation.  We analytically 
relate $L_\lambda$ to $L_{bol}$, assuming
purely blackbody emission, and that the bolometric mass-luminosity 
relation can be expressed as a power law:
\begin{equation} \label{eqn:ml}
L_{bol} = L_{{bol},\odot} \left(\frac{M}{M_\odot}\right)^{\beta},
\end{equation}
in which $\beta$ is a single number -- the bolometric power law index -- instead 
of the wavelength-dependent index $\beta_\lambda$ in \eq{eqn:mrl}.  
Empirically, this is known to be a reasonable approximation for $0.3M_\odot \la M \la 2 M_\odot$ \citep{popper80}.
We combine 
this bolometric relation with the mass-radius relation from \eq{eqn:mrl}, and 
with $L_{bol} = 4 \pi R^2 \sigma T^4$.  We can then write temperature as a 
function of mass,
\begin{equation}\label{eqn:TM}
T(M) = T_\odot \left(\frac{M}{M_\odot}\right)^{(\beta-2\alpha)/4}, \qquad
T_\odot=\left(\frac{L_{bol,\odot}}{4\pi\sigma R_\odot^2}\right)^{1/4},
\end{equation}
where $T_{\odot}= 5777~{\rm K}$ is the effective temperature of the sun.  We can 
write the luminosity of a blackbody in a particular band $X$ as:
\begin{equation} \label{eqn:LTl}
L_X(T) = \int^{+\infty}_{-\infty} \ct_X(\lambda^\prime)B_{\lambda^\prime}(T) (4 \pi R^2)(\pi) d\lambda^\prime,
\end{equation}
where $B_{\lambda}(T)$ is the Planck law per unit wavelength,
and $\ct_X(\lambda)$ is the transmission for filter X.  We can approximate 
this formula by assuming that the transmission $\ct_X(\lambda)$ 
of filter $X$ is a simple
top hat with unit height, effective width $\Delta \lambda_X$, 
and effective wavelength $\lambda_{c,X}$.  We can also 
replace the integral with a product, since $B_{\lambda}(T)$ does 
not change significantly over the intervals defined by the 
visible or near-infrared filters we will be considering.  
Also, we can use \eq{eqn:TM} to 
write $L_X$ as a function only of mass.  Thus, we can rewrite equation (\ref{eqn:LTl}) as
\begin{equation} \label{eqn:LTl2}
L_X(M) = \frac{8 \pi^2 c R^2 \lambda_{c,X}^{-4} \Delta\lambda_X}{\exp\left(\frac{hc}{\lambda_{c,X} k T(M)}\right) - 1} \, .
\end{equation}

To check this form of the luminosity function, we compare its reported 
luminosities to those from the Yale-Yonsei (${\rm Y}^2$) isochrones \citep{yi01}, which 
use the \citet{lej98} color calibration.  We find that this
form for $L_X(M)$ is sufficiently accurate for our purposes.  In
particular, it is much more accurate than the simple power-law
approximations we considered in \S\ref{sec:analapprox}.  Nevertheless,
we find that the qualitative conclusions outlined in that
section still holds using the more accurate form for the
mass-luminosity relation, and thus we can still use the intuition
gained by studying the behavior predicted by the analytic approximations
derived in \S\ref{sec:analapprox} to guide our interpretation of the results presented in the
rest of the paper.

\bigskip

\subsection{Normalizing the Mass Function} \label{sec:massfunc}

To normalize the mass function, we need to determine which values to
use for $M_{min}$ and $M_{max}$, which are used to compute 
the normalization constant $k$.  We should choose values that
limit the set of stars in the analysis to those around
which planets are likely to be detected. 

Somewhat anticipating
the results from the following sections, we will set $M_{min} = 0.3M_{\odot}$
for our fiducial calculations.  This 
represents the lowest mass star around which a planet
can be detected, for typical ranges of the observational,
system, and planet parameters encountered in current transit searches.  In some cases,
it may be possible to detect planets around stars of lower mass.  On the other
hand, one might be interested in only those stars for which precise radial velocity
follow-up is feasible for 8m-class telescopes.  Therefore, in \S\ref{sec:mmass},
we consider the effects of varying $M_{min}$ on the number of detectable
planets.  

We set $M_{max}$ to be the most
massive main sequence star in the system, i.e. a turnoff star.
We determine the mass of a turnoff star, $M_{to}$, using the simple relation
$L_{bol,to} = \epsilon M_{to} c^2 / A$, where $\epsilon$ is the net
efficiency of hydrogen burning ($\epsilon = 0.00067$) and $A$ is the
age of the target system.  Combining this expression with the bolometric
mass-luminosity relation from \eq{eqn:ml} gives us 
\begin{equation}
M_{to} = \left(\frac{\epsilon M_\odot^\beta c^2}{L_{bol,\odot} A}\right)^{1/(\beta-1)}.
\label{eqn:mto}
\end{equation}

\bigskip

\subsection{Minimum Observational Error} \label{sec:error}

In \S\ref{sec:psn}, we calculate
$\sigma$ using a formula for pure photon noise errors.  In real
observations, photometric errors do not get arbitrarily precise for a
given source and background.  Therefore, we impose a minimum
systematic observational error of $\sigma_{sys}=0.1\%$ to mimic the
practical difficulties of obtaining precise observations of bright
stars.  The calculated errors therefore become equal to
$\sigma=\left(\sigma_{phot}^2 + \sigma_{sys}^2\right)^{1/2}$, where
$\sigma_{phot}$ is the photon-noise error.

\bigskip

\subsection{Effective Area of the Seeing Disk}

We assume the point-spread function (PSF) is a Gaussian with
a full-width half-maximum of $\theta_{see}$, which
has an effective area of,
\begin{equation}\label{eqn:aeff}
\Omega= \frac{\pi}{\ln 4} \theta_{see}^2.
\end{equation}

\bigskip

\subsection{Saturation Mass}

Detectors have a finite dynamic range, and we clearly
cannot detect planets around saturated stars.  When integrating over
mass, we therefore ignore stars with $M\ge M_{sat}$, where
$M_{sat}$ is the mass of a star that just saturates
the detector.  We assume that a star saturates 
the detector when the number of photons $N_{phot}$ 
from the star and sky that fall into the central pixel of the stellar PSF 
exceeds the full well depth of a pixel, $N_{FW}$.  
We approximate $N_{phot}$ as,
\begin{equation} \label{eqn:Nphot}
N_{phot} = \left( f_\lambda \left\{ 1 - \exp \left[ - \ln 2 \left(\frac{\theta_{pix}}{\theta_{see}}\right)^2 \right] \right\} + S_{sky,\lambda} \theta_{pix}^2\right)t_{exp} \pi \left(\frac{D}{2}\right)^2 ,
\end{equation}
where $\theta_{pix}$ is the angular size of a single pixel.  This form 
assumes a Gaussian PSF perfectly
centered on the central pixel.  The assumption of Gaussian PSF
is reasonable for our purposes, and the assumption that the PSF is centered on
a pixel conservatively underestimates $M_{sat}$.  Formally,
\eq{eqn:Nphot} only holds for circular pixels, but is nevertheless
accurate to $\la 20\%$ for square pixels.  This is sufficient for
our purposes. 

\bigskip

\subsection{Planet Distribution} \label{sec:PlanDist}

To compute the number of detected planets, we 
integrate $d^3N_{det}/drdPdM$ (see \eq{eqn:dN}) over $M$, $P$, and $r$ to find $N_{det}$. We therefore 
must assume a form for the distribution of planets, $d^2p/drdP$.
We will assume that the periods are  
distributed evenly in log space, as is suggested by several
analyses (e.g., \citealt{tt02}).  Since radii have been measured for only seven
planets, the distribution of the radii is very poorly
known.  We will therefore simply assume a delta function 
at $r=r'$, and adopt $r'= 0.1 R_{\odot}$ 
for our fiducial calculations.  However, we will also
explore the detectability as a function of $r$.  Our adopted
distribution of periods and radii can therefore be expressed as,
\begin{equation}\label{eqn:dpdrdp}
\frac{d^2p}{drdP} = \frac{1}{\Delta \ln P} P^{-1} \delta(r-r') \,,
\end{equation}
where $\Delta \ln P$ is the logarithmic range of periods of interest.

From a comparison of the results from radial velocity and transit surveys,
it appears that there are two distinct populations of close-in massive planets.
``Very Hot Jupiters'' have periods between $1-3~{\rm days}$, and are
approximately ten times less common than ``Hot Jupiters'' with
periods between $3-9~{\rm days}$ \citep{gsm05}.  We will therefore
consider these two ranges of periods separately. 

\bigskip

\subsection{Extinction}

We consider two models for the extinction.  In general, we assume an
extinction of a fixed value $A_I$ in the $I$-band, and calculate the
extinction in the other bands using the extinction ratios listed in
Table \ref{tab:mags}.  We also consider an extinction that depends on
the distance to the stellar system as,
\begin{equation}
A_I(d) = 0.5~{\rm mag} \left(\frac{d}{\rm kpc}\right) \, ,
\label{eqn:aidist}
\end{equation}
where we again use the the extinction ratios in Table \ref{tab:mags}
to determine the extinction in the other bands.  We use the
fixed-extinction law in all calculations and plots unless otherwise
specified.

\bigskip

\subsection{Fiducial and Fixed Parameters}

There are a number of parameters in these equations for which we must
assign values.  In \S\ref{sec:ObsParams} we will examine the
dependence of the detection probabilities on a subset of the most
interesting of these parameters.  These include the cluster distance
$d$, age $A$, mass function slope $\gamma$, and extinction $A_I$, as
well as the telescope aperture $D$, the exposure time $t_{exp}$,
the seeing $\theta_{see}$, duration of the survey $N_n$, detection
threshold $\Delta\chi^2_{min}$, and planet radius $r$ and orbital
period $P$.  Our choices for the fiducial values of these parameters
are listed in Table \ref{tab:fids}.  We do not vary the values of the
other parameters, either because they are quantities that are
empirically well-determined, or because their values are specific to
the kinds of surveys we are considering here.  These quantities are
the detector readout time $t_{read} = 30~{\rm seconds}$, 
the fullwell depth of
the detector $N_{FW} = 10^5$ photons, and the angular size of the
detector pixels $\theta_{pix} = 0.2$ arcsec.  We assume an
exponent of the mass-radius relation of $\alpha = 1$.  We also assume
that observations can take place during 7.2 hours each night, and we
require two transits to be observed for a detection.  The fiducial 
values chosen in Table \ref{tab:fids} are not intended to represent a 
specific cluster, but rather to be typical values for star clusters in the Galaxy.

\begin{deluxetable}{cc}
\tablecaption{\sc Fiducial Parameters}
\tablewidth{0pt}
\tabletypesize{\scriptsize}
\tablehead{
  \colhead{Parameter} &
  \colhead{Value}}
\startdata
Distance ($d$) & 2.5 kpc \\ 
Age ($A$) & 1 Gyr \\ 
Mass Function Slope ($\gamma$) & -2.35 \\
Bolometric Index ($\beta$) & 4.0 \\
Extinction in I-band ($A_I$) & 1.25 \\
Telescope Aperture ($D$) & 200 cm \\ 
Exposure Time ($t_{exp}$) & 60 s \\ 
Seeing ($\theta_{see}$) & 1 arcsec \\ 
Chi-Square Threshold ($\Delta\chi^2_{min}$) & 30 \\
Duration of Survey ($N_n$) & 20 nights \\
Planet Radius ($r$) & 0.1 $R_{\odot}$ \\
Orbital Period ($P$) & 2.5 days 
\enddata
\label{tab:fids}
\end{deluxetable}

Since $\psn$ depends on the observational band, we calculate $\psn$
(and by extension, the overall probability $\ptot$) using 4 different
bands, $I$, $V$, $B$, and $K$, using the $\lambda_{c,X}$ and $\Delta
\lambda_X$ for each band as defined in \citet{bessell98}.   We use the sky brightness in the
different bands, $\mu_I$, $\mu_V$, $\mu_B$, $\mu_K$, from the KPNO
website.\footnote{http://www.noao.edu/kpno/manuals/dim/dim.html\#ccdtime}
Table \ref{tab:mags} lists the values for sky brightness, along with
the flux zero point values, which come from \citet{bessell98}.

\begin{deluxetable}{cc}
\tablecaption{Sky brightnesses, Zero-point Fluxes, and Extinction Ratios}
\tablewidth{0pt}
\tabletypesize{\scriptsize}
\tablehead{
  \colhead{Sky brightness\tablenotemark{a}} &
  \colhead{magnitude per ${\rm arcsec}^{-2}$ }}
\startdata
$\mu_I$ & 20.0 \\ 
$\mu_V$ & 21.8 \\ 
$\mu_B$ & 22.7 \\
$\mu_K$ & 13 \\ 
\hline
Zero point flux\tablenotemark{b} & W ${\rm cm}^{-2} \mu {\rm m}^{-1}$ \\ 
\hline
$f_{0,I}$ & $1.13 \times 10^{-12}$ \\ 
$f_{0,V}$ & $3.63 \times 10^{-12}$ \\ 
$f_{0,B}$ & $6.32 \times 10^{-12}$ \\ 
$f_{0,K}$ & $3.96 \times 10^{-14}$ \\ 
\hline
Extinction ratio\tablenotemark{c} &  \\ 
\hline
$A_V / A_I$ & 2.07 \\
$A_B / A_I$ & 2.74 \\
$A_K / A_I$ & 0.232
\enddata
\tablenotetext{a}{from the KPNO website, http://www.noao.edu/kpno/manuals/dim/dim.html\#ccdtime}
\tablenotetext{b}{from \citet{bessell98}}
\tablenotetext{c}{from \citet{bm98}, recalculated for this paper using $I$ as the reference band.}
\label{tab:mags}
\end{deluxetable}

\bigskip

\section{Results}\label{sec:results}

We now have all the pieces we need to use \eq{eqn:dN} to evaluate the
number of planets $N_{det}$ that can be detected by a particular
survey toward a given stellar system.  Our objective in this section
is to explore how the overall detection probability depends on the
various properties of the stellar system, the planets, and the survey,
and to provide an estimate of the yield of planets for a particular
transit survey.  

We begin by exploring the detection sensitivity as a function of host
star mass, confirming the basic conclusions we derived from our
simple analytic considerations presented in \S\ref{sec:analapprox}.
We then consider the detection probability as a function of period,
integrated over the mass function of the system.  Finally, we consider
the fraction of detected planets as a function of the various
observational and cluster properties, fully integrated over the mass
function, as well as the assumed planetary period distribution.
Unless otherwise stated, we will adopt the fiducial assumptions and
parameter values described in detail in \S\ref{sec:AddIngr}.

\bigskip

\subsection{Sensitivity to Host Star Mass $M$\label{sec:hsres}}

\begin{figure}[t]
\epsscale{1.0}
\plotone{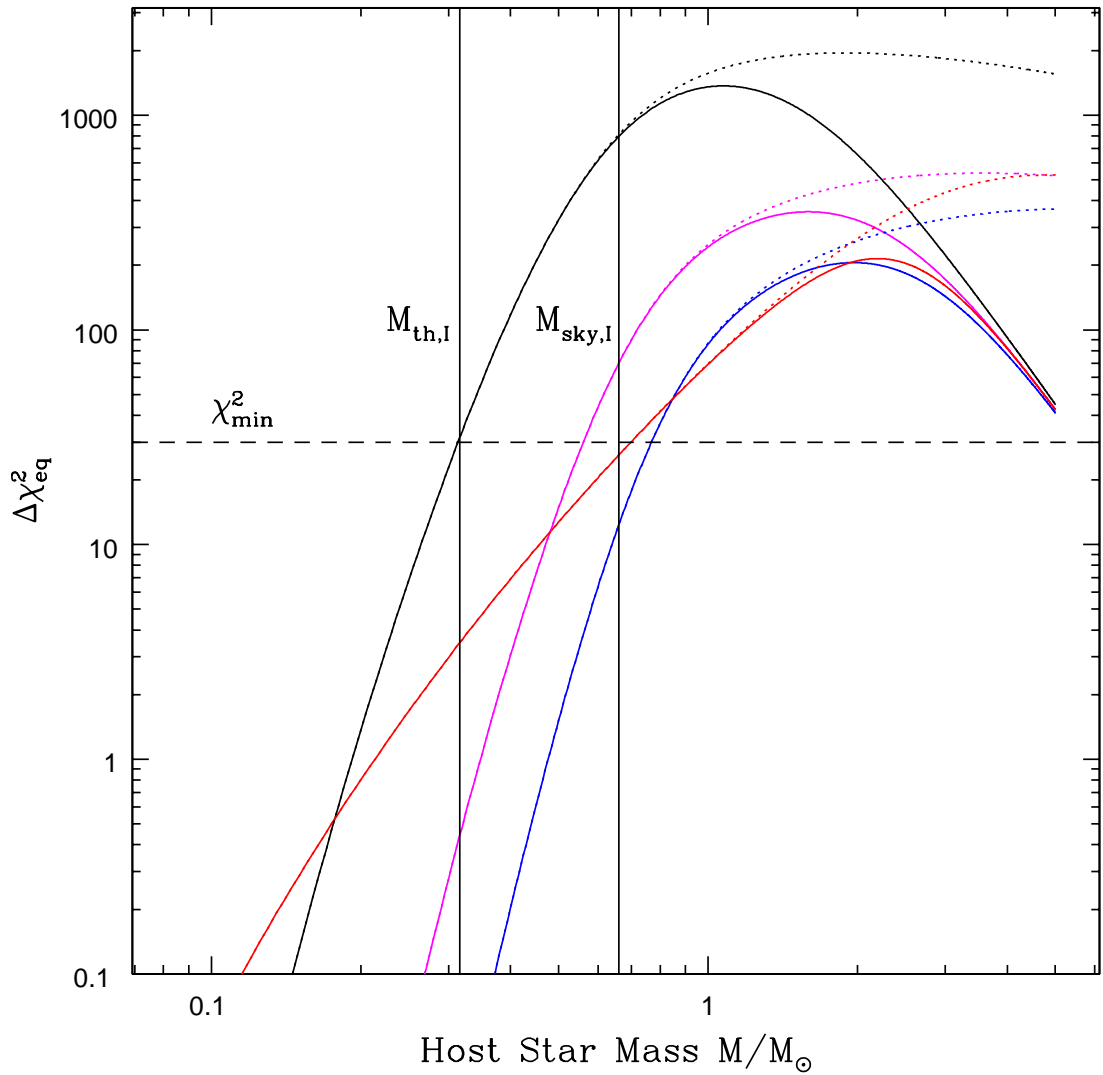}
\caption{$\Delta\chi^2_{eq}$ versus host star mass for our
fiducial parameters and $r=0.1R_\odot$ and $P=2.5~{\rm days}$.  The vertical lines show $M_{{th},I}$ and
$M_{sky,I}$, while the horizontal dashed line shows our fiducial value for $\Delta\chi^2_{min}$.  The dotted lines show the curves of $\Delta\chi^2_{eq}$ without the inclusion of the systematic error $\sigma_{sys}=0.1\%$.}
\label{fig:chi}
\end{figure}

We first consider the sensitivity as a function of the host star mass.
We begin by considering $\Delta\chi^2_{eq}$ versus mass for our
fiducial parameters and $r=0.1R_\odot$ and $P=2.5~{\rm days}$.  This
is shown in Figure \ref{fig:chi}, for the four photometric bands we 
consider.  We also show our fiducial value of
$\Delta\chi^2_{min}=30$, and the quantities $M_{{th},I}$ and
$M_{sky,I}$ introduced in \S\ref{sec:psn_on_M}.  In order to
elucidate the effects of systematic errors, we show
$\Delta\chi^2_{eq}$ for our no systematic error,
and for our fiducial assumption of a systematic
error of $\sigma_{sys}=0.1\%$.  When the
systematic error is negligible, we find that $\Delta\chi^2_{eq}$
is approximately independent of mass for $M\ga M_{sky}$, as
anticipated in \S\ref{sec:psn_on_M}.  However, when systematic errors
are included, $\Delta\chi^2_{eq}$ has a peak, which for our
adopted values is near $M_\odot$.  We also see that, for all of the
photometric bandpasses and fiducial parameter values we consider, the surveys are in the
background-limited regime, and that the S/N is highest in the
$I$-band, implying that, all else equal, the number of detected
planets will be maximized when using this band for these
fiducial parameter values.

It is interesting to note in Figure \ref{fig:chi} that the behavior
of $\Delta\chi^2_{eq}$ versus mass is fundamentally different in $K$ than
the optical bandpasses.  The basic reason for this
is that, for the mass range considered here ($0.1 \la M/M_\odot \la 2$),
observations in $K$ sample the stellar spectrum in
the Rayleigh-Jeans tail, whereas observations in the optical
sample near the blackbody peak or in the Wein exponential tail.  Therefore,
the $\Delta\chi^2_{eq}$ falls more gradually toward lower masses
for observations in $K$.  We will see this fundamentally different behavior
in $K$ exhibited in many of the following results. 

\begin{figure}[t]
\epsscale{1.0}
\plotone{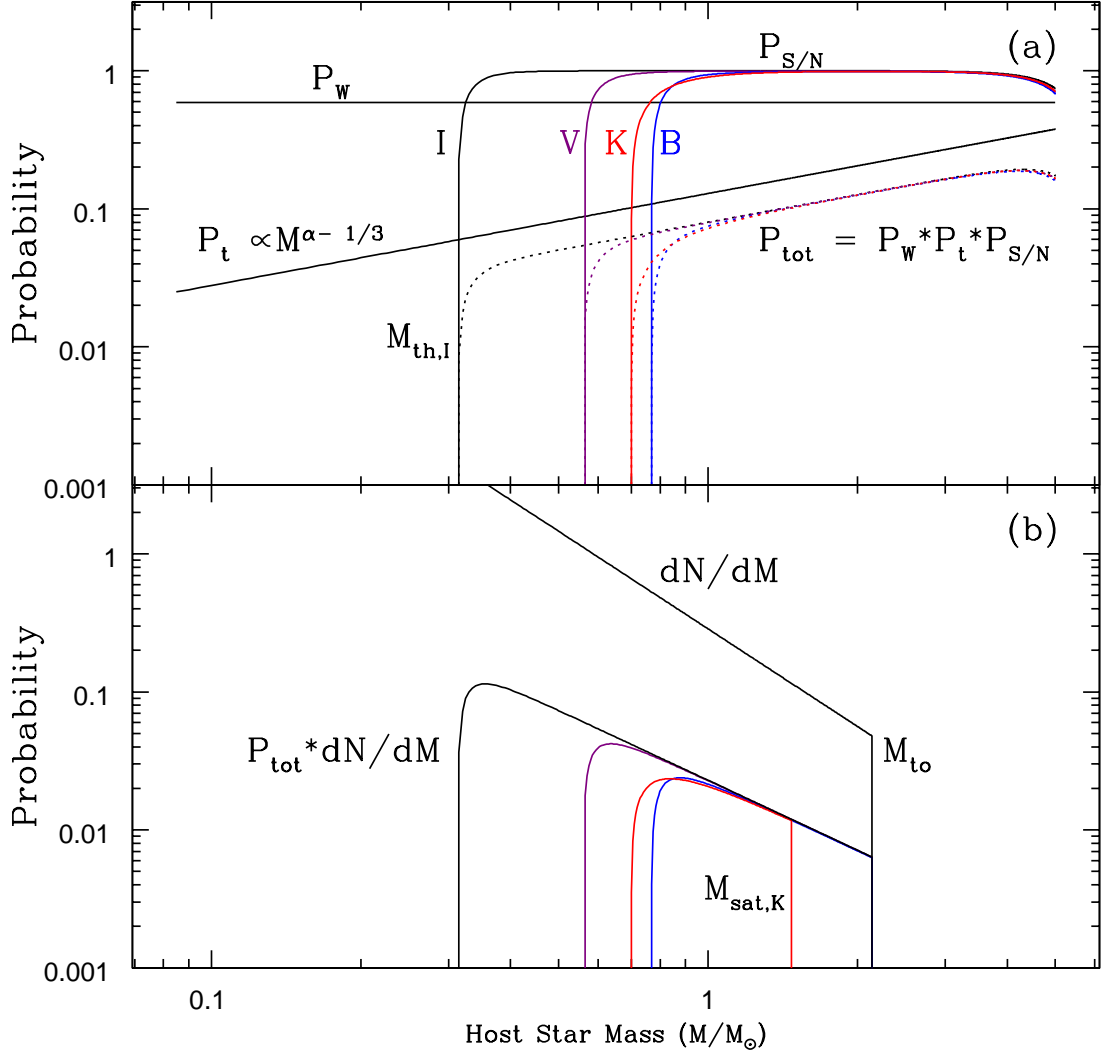}
\caption{
Detection probabilities versus mass.  Panel (a) shows $\pw$, $\pt$,
and $\psn$, and the product of all three, $\ptot$.  The curves 
for $\ptot$ are displayed with dotted lines, to make them more 
easily distinguishable from the curves for $\psn$.  Panel (b)
shows the mass function $dN/dM$, along with the product of it and
$\ptot$.  In both plots, $\psn$ and $\ptot$ are shown in four
different bands, $I$ (black), $V$ (purple), $B$ (blue), and $K$ (red).
The plots are calculated using the fiducial parameter values
in Table \ref{tab:fids}, and a period of $P = 2.5$ days, for which
$\pw = 0.63$.  The cutoff in (b) is due to the fact that
for a system with the fiducial parameter values used for this plot,
the turnoff mass is a little over $2 M_{\odot}$.}
\label{fig:main}
\end{figure}

We next consider the overall detection probability $\ptot(M,P,r)$, and
its various components, $\pw$, $\pt$, and $\psn$.
$\pt$ is described by equations
(\ref{eqn:ptmass}) and (\ref{eqn:c3}); $\psn$ is described by
equations (\ref{eqn:generic_psn}), (\ref{eqn:c1}), and (\ref{eqn:c2});
and the window function $\pw$ is shown in Figure (\ref{fig:win}).  We
plot these various detection probabilities, and
$\ptot$, versus host star mass for our set of fiducial parameter
values in Figure \ref{fig:main}a.  The overall shape of the
$\psn$ curve in the top plot is simple; it shows that a transit will
be detected if the star's mass is greater than $M_{th}$.  The
small downturn at high masses is due to the systematic error
introduced in \S \ref{sec:error}.  That is because as $M$ increases, the depth of the
transit decreases (because of the increasing $R$) but the photometric
precision also increases.  However, by placing a limit on the measured
precision, at high masses the decrease in $\delta$ is no longer offset
by a decrease in $\sigma$, and so the sensitivity dips.

We combine these pieces with the mass function $dn/dM$ in Figure
\ref{fig:main}b.  There are a couple interesting features of the
lower plot.  The mass function cuts off a little over $2M_\odot$
because that is the turnoff mass for a system with the fiducial
parameters we are using.  The probability curve for $K$ band cuts off
before that point, though.  That is because a detector with the
fiducial values we have chosen ($D = 2$m, $t_{exp} = 60$s,
$\theta_{pix} = 0.2$ arcsec, and $N_{FW} = 10^5$ photons)
saturates at that mass in $K$, while the values for $M_{sat}$ for
$I$, $V$, and $B$ are higher than $M_{to}$ for this fiducial
stellar system.

Looking at the lower plot, it is clear that $I$ band is the best one
to use to detect planets.  The number of stars increases with decreasing
mass, and it is possible to detect planets around stars of lower mass
in the $I$-band.  Since $N_{det}$ involves the integral of
$\ptot(dn/dM)$ over mass, the total number of planets detected will be larger
in the $I$-band.  We will see in \S\ref{sec:ObsParams} that the $I$
band remains optimal for most parameter combinations encountered in
current transit surveys.

\bigskip

\subsection{Sensitivity to Period}

\begin{figure}[t]
\epsscale{1.0}
\plotone{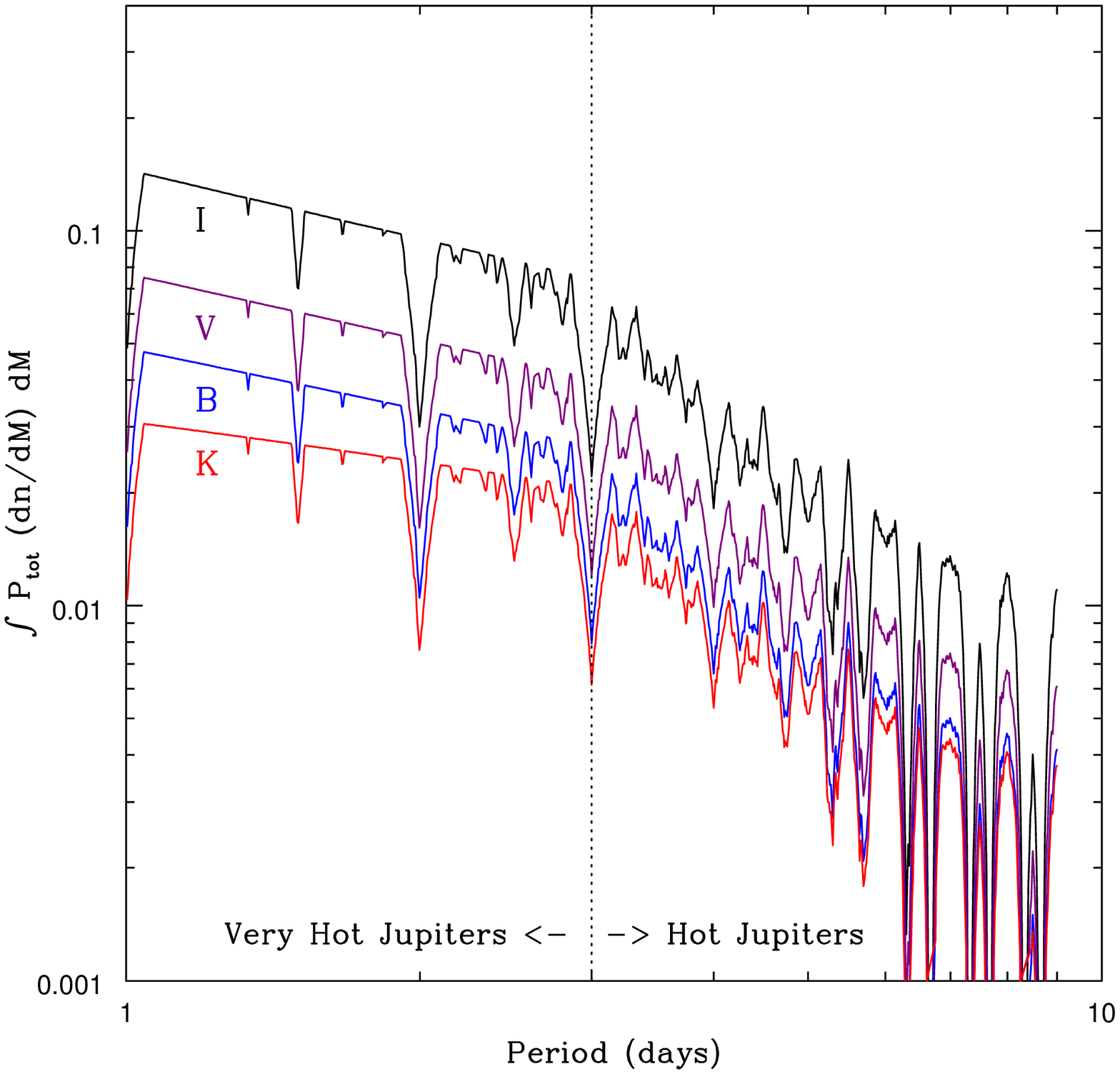}
\caption{
The total detection probability $\ptot$, weighted by the mass function
$dn/dM$, and integrated over mass, i.e.\ $\int \ptot (dn/dM) dM$,
versus planet period, for our fiducial parameters (see Table
\ref{tab:fids}). The window function used for this calculation is the
same $\pw$ as in Figure \ref{fig:win}.  The various colors represent
the different observing bands, $I$ (black), $V$ (purple), $B$ (blue),
and $K$ (red).}
\label{fig:per}
\end{figure}

We next examine the sensitivity as a function of period.  We consider
the total detection probability $\ptot$, weighted by the mass
function, $dn/dM$, integrated over period, i.e.\ $\int \ptot (dn/dM)
dM$.  This is shown in Figure \ref{fig:per}.  

The strong sensitivity to shorter period planets is clear,
and arises from competition from several effects.  The signal-to-noise
probability  $\psn$ increases for increasing $P$, since a planet with a longer
period will have a transit with a longer duration, and so there will
be more observations during the transit and
hence higher S/N.  However, this effect is more than compensated by the 
fact that the transit probability is $\propto P^{-2/3}$, and the window
probability $\pw$ generally
increases for smaller periods (see Figure \ref{fig:win}), since there is
a greater chance of detecting two transits for shorter periods.

\bigskip

\subsection{Sensitivity to Parameters} \label{sec:ObsParams}

\begin{figure}[t]
\epsscale{1.0}
\plotone{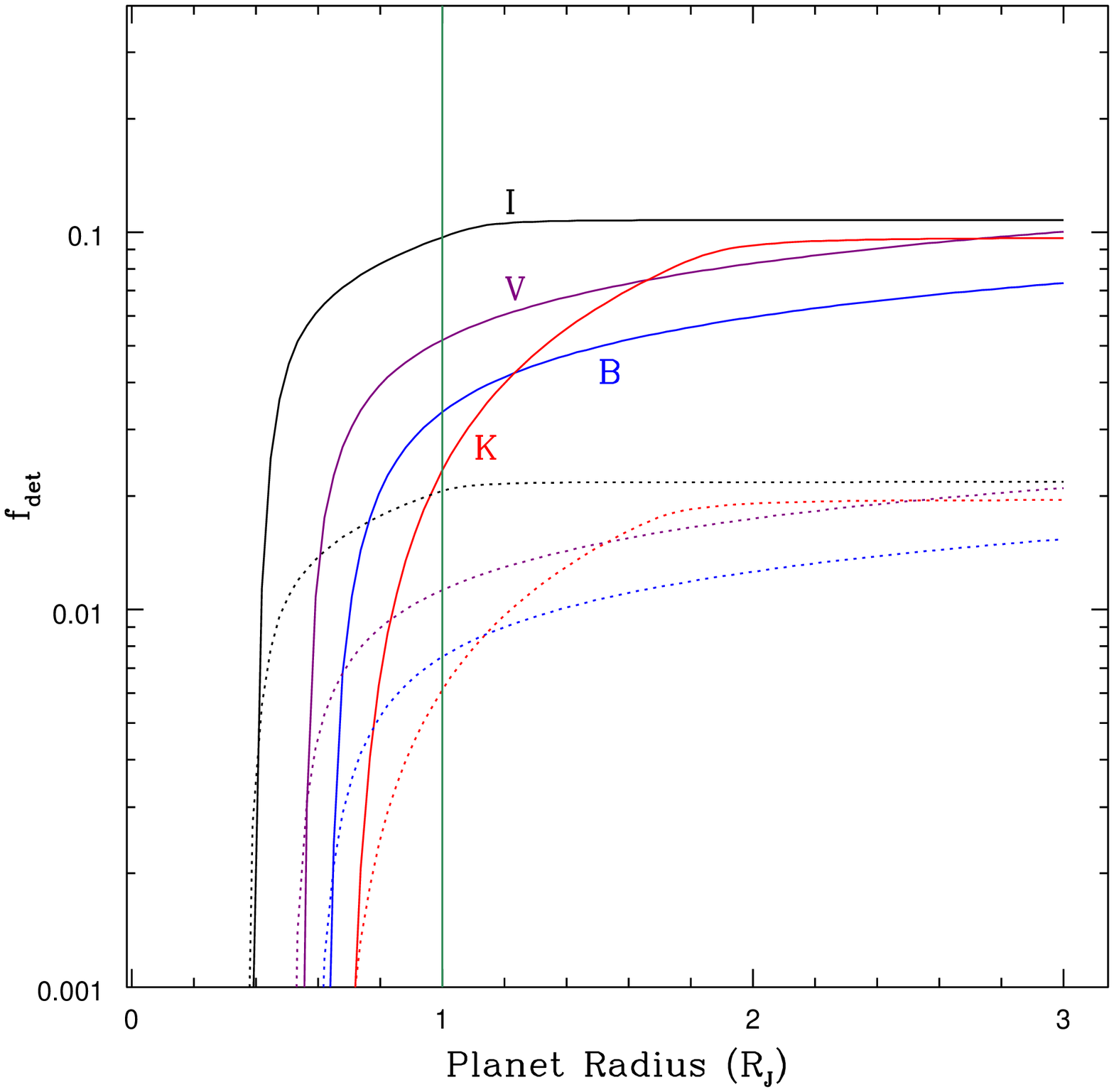}
\caption{The fraction of planets detected, $\fdet$ versus 
planet radius, for our fiducial parameters (see Table \ref{tab:fids}).  
In order to convert to the number of planets detected $N_{det}$, 
these numbers should be multiplied by the total number of stars $N_*$ in the system
with masses between $M=0.3M_\odot$ and the turn-off mass, and by the fraction $f_p$
of these stars with planets, i.e.\ $N_{det}=\fdet f_p N_*$.
The colors represent the different observing 
bands, $I$ (black), $V$ (purple), $B$ (blue), and $K$ (red).  The solid lines 
show $\fdet$ for ``Very Hot Jupiters,'' i.e.\ planets
with periods $P=1-3~{\rm days}$.  The dotted lines 
show $\fdet$ for ``Hot Jupiters,'' planets
with periods $P=3-9~{\rm days}$.  The vertical green line indicates the fiducial value we use 
for the other plots, $r = 0.1R_{\odot} \simeq 1R_J$.}
\label{fig:rad}
\end{figure}

In this section, we examine how the fraction of detected
planets $\fdet$ depends on the various input parameters considered and
listed in Table \ref{tab:fids}.  Conceptually, there are three 
different classes of parameters in Table \ref{tab:fids}.  Five of
the parameters describe the properties of the target 
system: $d$, $A$, $\gamma$, $\beta$, and $A_I$.  Five of the
parameters are properties of the observing setup: $D$, $t_{exp}$,
$\theta_{see}$, $\Delta\chi^2_{min}$, and $N_n$.  The two remaining parameters, $P$ and $r$, are properties of
individual planets.

Integrating over mass, period, and radius, the fraction
of planets detected is,
\begin{equation}
f_{det}\equiv \frac{N_d}{f_p N_*}=\int \int \int  dr dP dM \frac{d^2p}{drdP} \ptot(M,P,R) \frac{dn}{dM} \, .
\label{eqn:fdet}
\end{equation}  

\begin{figure}[t]
\epsscale{1.0}
\plotone{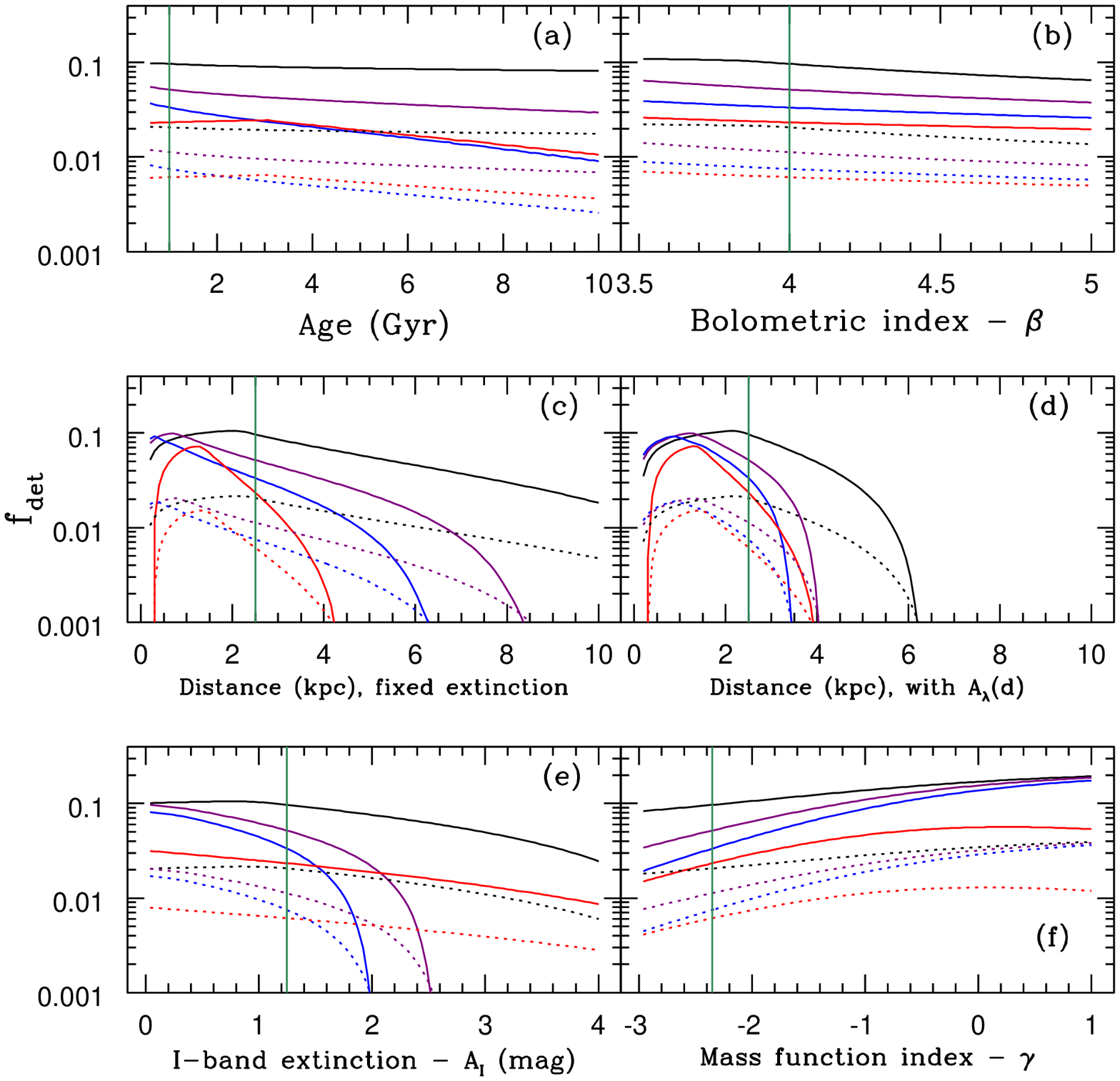}
\caption{The fraction of planets
detected, $\fdet$, as a function of the parameters of the target 
stellar system.  
Line types are the same as in Figure \ref{fig:rad}. 
The vertical green line in each
plot indicates the fiducial value for that parameter used in calculating 
all the other plots.  The panels show $\fdet$ versus (a) age of the system,
(b) power-law index of the mass-bolometric luminosity relation, (c) distance 
to the system, for a fixed extinction, (d) distance to the system, for a 
distance-dependent extinction, (e) $I$-band extinction, and (f) the index of the mass function.}
\label{fig:params_c}
\end{figure}

\begin{figure}[t]
\epsscale{1.0}
\plotone{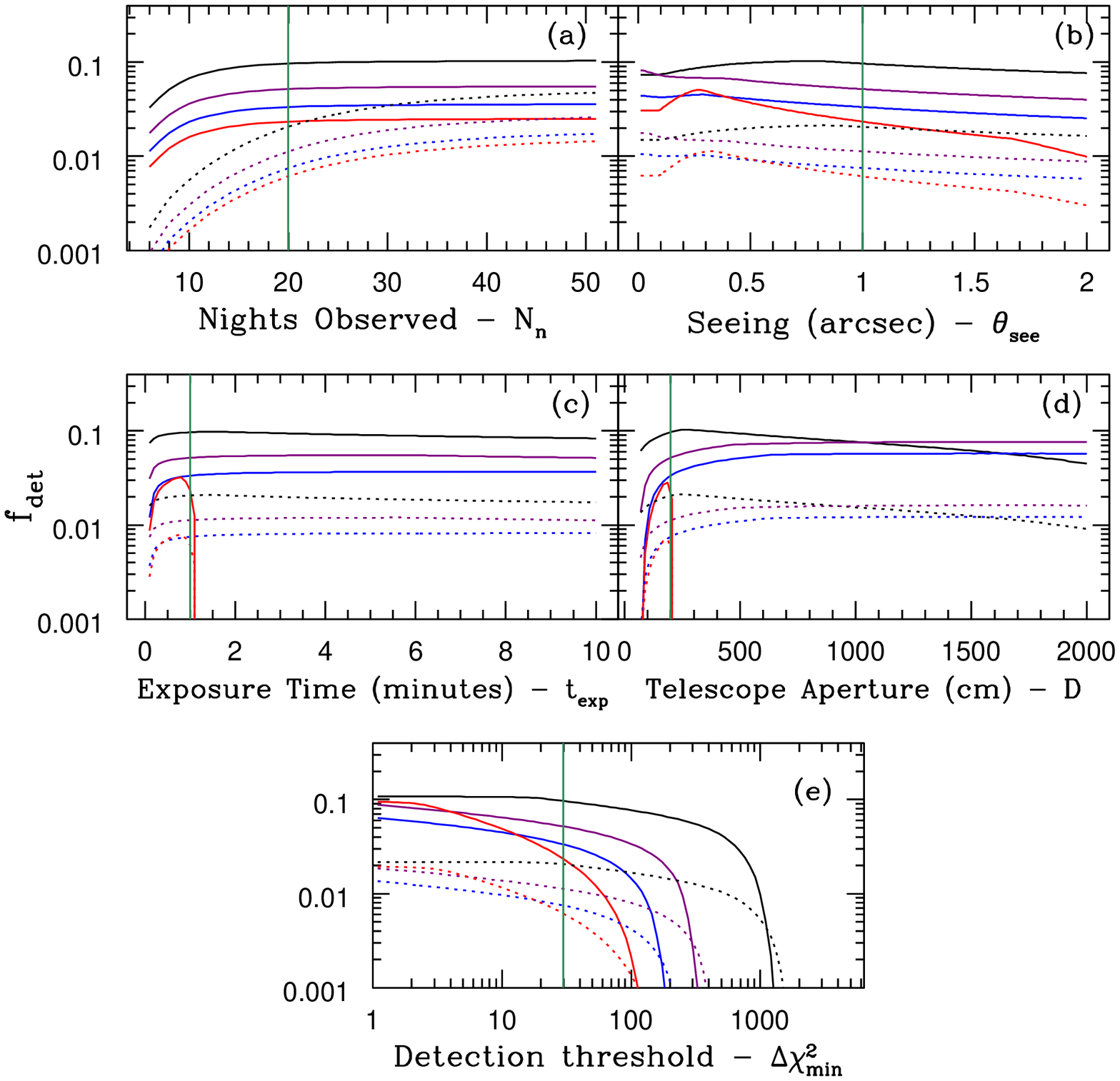}
\caption{Same as Fig.\ \ref{fig:params_c}, but with $f_{det}$ as a function
of the observational parameters.    The panels show
$\fdet$ versus (a) number of nights of the observational
campaign,
(b) full-width half-maximum of the point-spread function (the seeing),
(c) the exposure time, (d) the diameter of the telescope, and (e) 
the minimum $\chi^2_{min}$ required for detection. 
}
\label{fig:params_o}
\end{figure}

In Figure \ref{fig:rad}, we plot $\fdet$ versus planet radius.
We plot $\fdet$ for the six parameters of the target system 
in Figure \ref{fig:params_c}, and the five observing parameters 
in Figure \ref{fig:params_o}.  
We shall now go though each of the parameters and describe the dependencies.

\begin{itemize}

\item {\bf Radius} - The dependence on planetary radius $r$ shows that
detection probabilities increase very quickly up to the fiducial value
of $r = 0.1R_{\odot}$, at which point a transit with a planet of that
radius has sufficient signal to be detected around nearly all the
stars in the system.  In this plot we also see that while the curves
for $I$, $V$, and $B$ bands all have this similar ``step-function'' shape, 
whereas the rise for the $K$-band is more gradual.  

\item {\bf Age} - From Figure \ref{fig:params_c}a, we see $\fdet$ is
quite insensitive to the age of the system in $I$ band, and is
somewhat more sensitive in the other bands.  This is because, as we
see in Figure \ref{fig:main}b, a larger proportion of the planets
detected in the other bands are at higher masses.  The fact that
$M_{sat}$ is lower than $M_{to}$ in $K$ accounts for the break in
the curve in $K$.

\item {\bf Bolometric Index} - Figure \ref{fig:params_c}b shows that
$\fdet$ is weakly dependent on the value of $\beta$.  Therefore our
choice of $\beta = 4.0$ is not so important, as $\fdet$ is essentially
the same for $3.5 < \beta < 5.0$, which encompasses the whole range of
values that are typically used for the mass-bolometric
luminosity relation.

\item {\bf Distance} - This is a key parameter.  A nearby system will
have many saturated stars, which accounts for the turnover at small
distances.  For fixed extinction (Figure \ref{fig:params_c}c), as the
system gets further away, the signal drops and planets cannot be
detected around the smaller stars in the system. For
distance-dependent extinction (Figure \ref{fig:params_c}d), that
effect is compounded at large distances, although the extinction has
much less of an effect in $K$.  In both cases, for sufficiently
large distances, the system
transitions into the source-limited regime, at which point $\fdet$ drops
precipitously.  

\item {\bf Extinction} - Figure \ref{fig:params_c}e plots the
dependence on the value of $A_{I}$, showing the effects of greater
extinction at a fixed distance.  In a sense, combining the effects
from Figure \ref{fig:params_c}c and Figure \ref{fig:params_c}e gives
us Figure \ref{fig:params_c}d, although the combination is more
complex than a simple multiplication.

\item {\bf Mass Function} - The slope of the mass function $\gamma$
determines the relative number of smaller stars and larger stars.  The
fiducial value of $\gamma=-2.35$ is the usual Salpeter slope
\citep{salpeter55}.  As we see in Figure \ref{fig:params_c}f, the
detection probabilities do not depend greatly on the exact value of
the slope, although for larger values of $\gamma$ bluer bandpasses become more
competitive, as expected.

\item {\bf Nights Observed} - In Figure \ref{fig:params_o}a, we see
that the an observing campaign lasting about 15 nights will detect two
or more transits from nearly all the ``Very Hot Jupiters'' ($1~{\rm
day}< P < 3~{\rm days}$) that satisfy the detection threshold, but to
detect two or more transits from most of the detectable ``Hot Jupiters''
($3~{\rm days}< P < 9~{\rm days}$), the survey should last more than
twice as long.  Since we assume perfect weather in this analysis, even
more time should be expected to fully detect the most possible
transits.

\item {\bf Seeing} - An increase in the seeing means an increase in
the size of the PSF, and so an increase the number of pixels over which
the flux of the stars is distributed.  This affects $\fdet$ is two
distinct and opposite ways.  First, this increases the contribution of
the background noise at fixed mass, therefore increasing $M_{th}$.
Second, this decreases the number of photons in the central pixel, and
so increases the mass at which the detector saturates, $M_{sat}$.  As
discussed in \S\ref{sec:psn_on_M}, $M_{th}$ is rather weakly
dependent on seeing for experiments in the background limited regime,
due primarily to the fact that the mass-luminosity relation is so
steep.  Furthermore, the increase in $M_{th}$ is partially
compensated for by the increase in $M_{sat}$.  As a result, the
$\fdet$ varies very little for the typical range of seeing encountered
in real observations, as seen in Figure \ref{fig:params_o}b.  Since
the sky is so much brighter in $K$, the seeing dependence is somewhat
greater in that band.

\item {\bf Exposure Time} - There are three effects of $t_{exp}$.
First, a longer exposure time increases the total number of photons in a
single observation.  Second, longer exposure times decrease the number
of observations per transit.  These two effects effectively cancel
when $t_{exp} \gg t_{read}$.  The third effect of $t_{exp}$
is that very long exposure times cause bright stars to saturate.  In
Figure \ref{fig:params_o}c the saturation effect is the reason why
$\fdet$ in $K$ falls so quickly, since for our fiducial setup the
large number of sky photons alone already brings the pixels close to
saturation in $K$.  In the other bands complete saturation does not occur even
at $t_{exp} = 10$ minutes.  Since saturation involves
both $t_{exp}$ and $D$, we shall see in \S \ref{sec:app} that the
simultaneous consideration of both factors is important.  

\item {\bf Telescope Aperture} - This factor enters in two ways.
Larger apertures allow a survey to reach the detection threshold for
fainter stars, yet also lead to saturation of brighter stars.  In
Figure \ref{fig:params_o}d in $K$ we see that $\fdet$ plummets a
little past $D = 200$~cm, since at that aperture the sky photons
alone saturate the pixels.  In $I$, the situation is complicated.
Increasing $D$ decreases $M_{th}$.  However, looking back at
Figure \ref{fig:main}, we see that in $I$, $M_{th}$ is just a
little larger than $0.3M_{\odot}$, which we take as the minimum
observable mass.  Thus increasing $D$ eventually pushes $M_{th}$
below $0.3M_{\odot}$.  Any further increase in $D$ will therefore not
result in additional detections at the low-mass end, and instead
simply lowers $\fdet$ as an increasing number of high-mass stars
saturate the detector. In $V$ and $B$, $M_{th}$ continues
to decrease for larger apertures, but due to its weak dependence on $D$, 
it is always above $0.3M_{\odot}$ for $D\le 2000~{\rm cm}$.
Further, the increase in $\fdet$ due to decreasing $M_{th}$ is compensated
for by the decrease in $M_{sat}$, such that
$\fdet$  is nearly independent of the aperture for $D\ga 500~{\rm cm}$
and $B-$ and $V-$band observations.

\item {\bf Detection threshold} - The choice of $\Delta\chi^2_{\rm
min}$ is strongly related to how much follow-up time and resources are
available for confirming transit candidates.  Since false positives
are a big hurdle in confirming transits, it is best to choose a high
value for $\Delta\chi^2_{min}$.  As anticipated in
\S\ref{sec:psn_on_M}, and seen in Figure \ref{fig:params_o}e, the
dependence of $\fdet$ on $\Delta\chi^2_{min}$ is relatively weak
until the background-limited regime is reached, at which point $\fdet$
falls rapidly.  For our fiducial parameter values
which are representative of many open cluster surveys, rather
stringent detection criteria of $\Delta\chi^2_{min} \la 100$ can
be tolerated without an unacceptably large reduction in the
detection efficiency.

\end{itemize}

\bigskip

\subsection{Minimum Mass} \label{sec:mmass}

Up to this point, our results have been normalized such that $\fdet$ is
the fraction of the planets orbiting stars with masses between
$0.3~M_\odot\le M \le M_{to}$ that are detected.  We have not
considered masses below $0.3~M_\odot$.  The lower mass limit was
chosen because this is approximately the minimum mass around which a
planet was detectable in the $I$-band for our fiducial assumptions
(see Fig.\ \ref{fig:chi}).  Furthermore, it is also
approximately the completeness limit of the deepest mass function
determinations for rich old open clusters (e.g., \citealt{kalirai01}).

\begin{figure}[t]
\epsscale{1.0}
\plotone{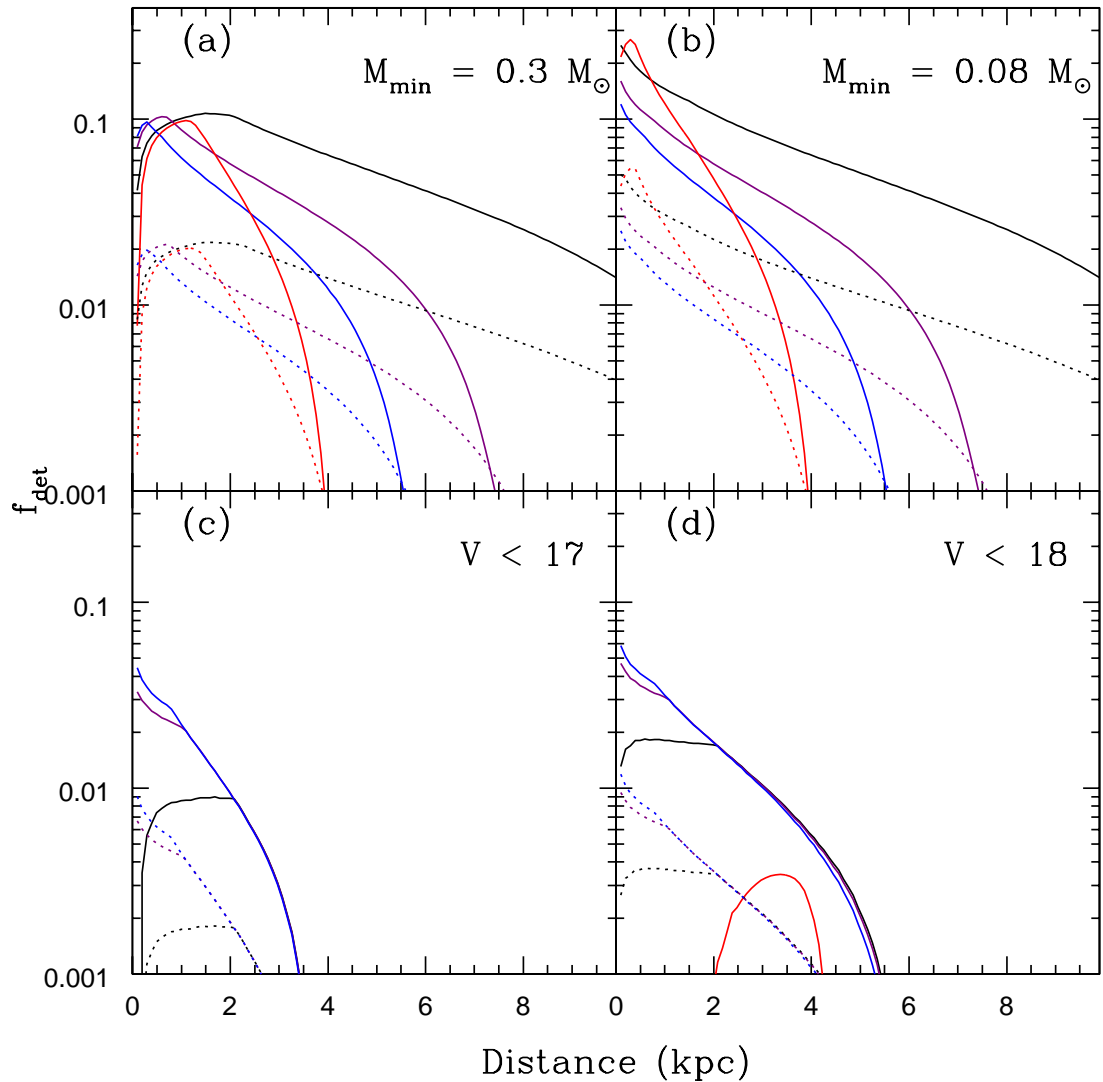}
\caption{Plots of $\fdet$ versus distance.  The different plots use
different criteria for the minimum mass cutoff, with a fixed mass of
$0.3M_{\odot}$ in Panel (a), $0.08M_{\odot}$ in Panel (b), or minimum masses corresponding to 
magnitude cuts $V < 17$ in Panel (c) and $V < 18$ in Panel (d).}
\label{fig:dist}
\end{figure}

In some instances, it may be possible to detect planets around stars
with masses considerably smaller than we have considered, with $M\le
0.3M_\odot$.  Since constraints on planets orbiting such very low mass
stars are meager, we briefly consider the detectability of planets
around host star masses in this regime.
Specifically, we perform the same analysis, except now we consider
stars with masses in the full range $M_{hb} \le M \le M_{to}$, where
$M_{hb}=0.08~M_\odot$ is the mass at the hydrogen-burning limit.  In
order to make these results directly comparable to our previous
results, we will continue to normalize the mass function such that
$N_*$ is the total number of stars between $0.3~M_\odot\le M \le
M_{to}$, and $N_{det}=f_p f_{det}N_*$, with $f_p$ the fraction of
stars with planets, and $f_{det}$ the number of planets orbiting
stars between $0.3~M_\odot\le M \le M_{to}$ that are detected. In
this way, $f_{det}$ can now formally exceed unity, although in
practice this is never the case.  Figure \ref{fig:dist}b shows $\fdet$
versus distance including stars down to the hydrogen-burning limit.
We see that, for monotonically rising mass functions, it may be
possible to increase the number of detections significantly by
considering very low mass primaries.  However, initial mass functions
are observed to have breaks near $M\sim (0.3-0.5)~M_\odot$, such that this
boost is probably not realized in practice, and furthermore any
detections around such low-mass primaries will be quite difficult to
confirm, as we discuss below.  Nevertheless, the potential for
constraining the planetary population of very low-mass primaries is
noteworthy.

In order to determine planet masses, as well as eliminate the many
kinds of astrophysical false positives that mimic planetary transits
\citep{torres04,mandushev05,pont05a}, reasonably precise $\sim 50~{\rm
m~s^{-1}}$ radial velocity (RV) follow-up measurements of candidate
transits are required.  Since the majority of the stars probed by
transit surveys toward stellar systems are relatively faint, the
ability to perform RV follow-up to this precision is a serious
concern.  The current state-of-the-art RV measurements on faint stars
using $10{\rm m}$-class telescopes can reach precisions of $\sim
50~{\rm m~s^{-1}}$ on stars with $V\la 17$
\citep{konacki03b,bouchy05,pont05b}.  It may be possible to push this
limit to somewhat fainter stars with more ambitious allocation of
resources, or improvements in future technology.  

In order to estimate what fraction of detected planets can be
confirmed using RV follow-up, in Figure \ref{fig:dist}, we plot
$\fdet$ versus distance, where we consider only those host stars with
apparent $V$-magnitudes of $V<17$ and $V<18$.  These can be compared
directly to the case where we consider all stars with $M\ge
0.3~M_\odot$.  The first feature of Figure \ref{fig:dist} that is
noticed is that the plots of $\fdet$ drop faster for the magnitude
limited cases than for the mass limited case, because of the nature of
the mass-luminosity relation.  For our fiducial parameters, it is
clear that the number of candidates that can be {\it confirmed} is
considerably smaller than the number that can be detected.
Furthermore, the advantage of observations in the $I$-band is
effectively removed, since most of the additional $I$-band detections
are too faint for RV confirmation.

It is clear that the ability to perform RV follow-up on candidate
planetary transits must be considered carefully when designing a
transit survey.  The question of how to devise a photometric survey
that maximizes the number of detected planets while accounting for the
ability to perform spectroscopic follow-up is outside of the scope of
this paper, but the formalism we have introduced here should provide the
tools to do so.
 
\bigskip

\section{An Application} \label{sec:app}

The most obvious application of our results is to use the predictions
for the number of detectable planets to choose optimal targets for a
particular survey, and to derive strategies to optimize the number of
detected transits for a specific target.  Since the specifics of the
optimal strategies will depend on detailed properties of the survey,
such as the site, detector, telescope, and time allotment, here we will
not attempt a comprehensive discussion, but rather simply suggest
heuristic guidelines motivated by a couple of specific examples.  

\begin{figure}[t]
\epsscale{1.0} \plotone{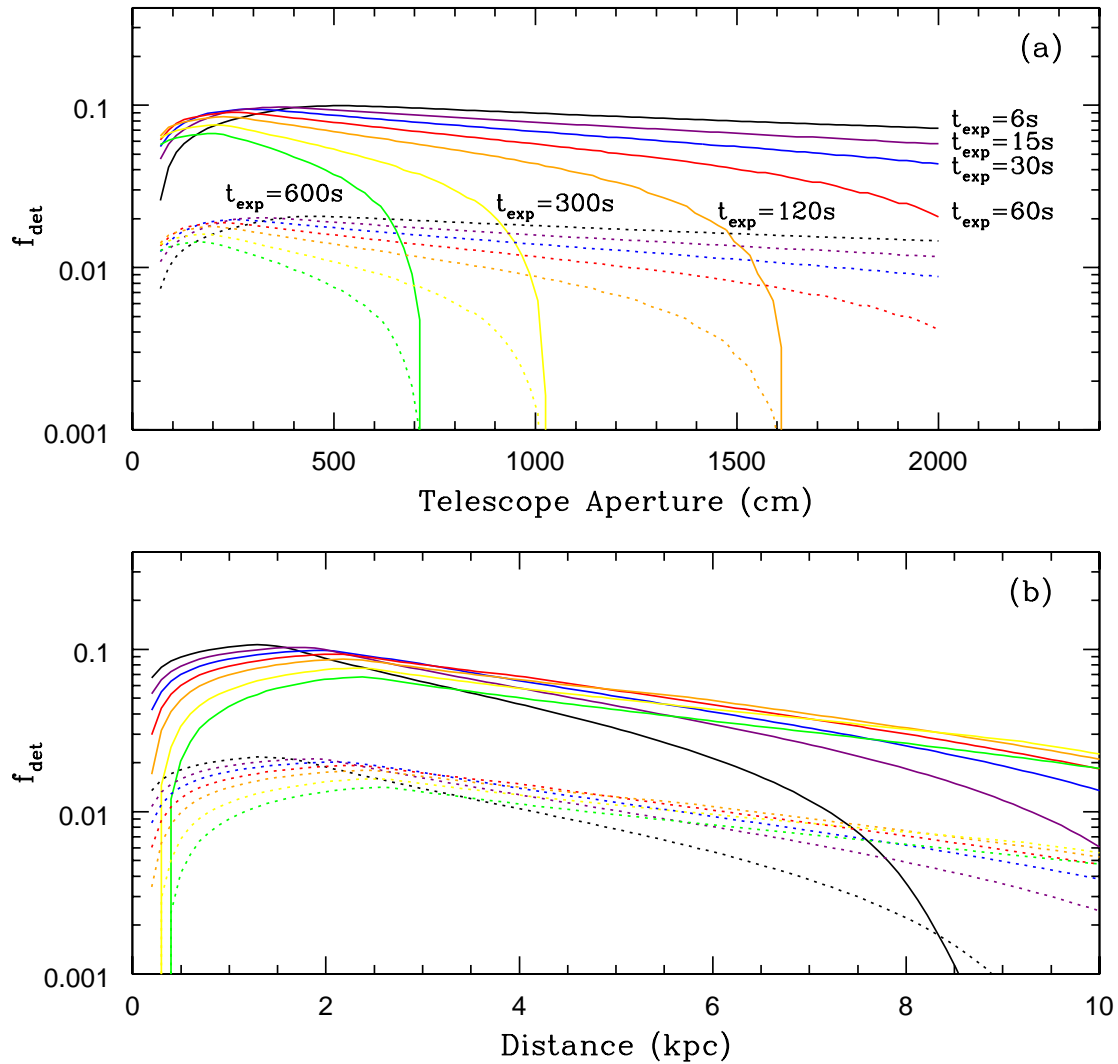}
\caption{The fraction of detected planets $\fdet$ as a function of
telescope aperture and distance for various exposure times, showing
only the I-band curves.  The solid lines represent Very Hot Jupiters
($1~{\rm day}< P < 3~{\rm days}$), while the dotted lines represent
Hot Jupiters ($3~{\rm days}< P < 9~{\rm days}$).}
\label{fig:opt}
\end{figure}

One important question is which targets are optimal in the sense of
allowing the largest number of possible detections.  There are a
number of factors that may enter into target selection, including
visibility, metallicity, richness, extinction, size and distance.
We illustrate how our results can be used to quantify and optimize
cluster selection, using the example of the trade-off between cluster
distance and exposure time.  As we have shown, $\fdet$ is a strong
decreasing function of cluster distance, such that closer clusters
are generally preferred.  However, saturation of bright stars is
also more problematic for more nearby clusters.  This can be partially
compensated for by decreasing the exposure time, but only until
$t_{exp}$ becomes comparable to $t_{read}$.  In Figure
\ref{fig:opt}b we show how $\fdet$ varies with distance to the target 
for various exposure times.  From this result, we see that 
clusters with distances $d\sim 2~{\rm kpc}$ are optimal, for telescope
apertures of $D\sim 200~{\rm cm}$. 

A somewhat different problem is to determine, given a particular
target system in which one wants to search for planets, what is the
optimal observational setup.  If the intent of a survey is to detect
all the transits of the brightest stars, then the exposure time should
be set such that the survey saturates at the turnoff stars of the
target system.  In that case, the exposure time can be 
calculated using equations (\ref{eqn:fnu}), (\ref{eqn:mrl}),
(\ref{eqn:Nphot}), and (\ref{eqn:mto}). On the other hand, if the
intent is to configure the parameters to achieve the largest number of
photometric detections, there are two factors at which we should
look more closely: aperture size and exposure time.  We can see in
Figure \ref{fig:opt}a how $\fdet$ varies with aperture size for
various exposure times.  When the aperture size is large, saturation
effects reduce the detection efficiency very quickly, and this can only be
partially compensated for by decreasing the exposure time.  As a
result, for transit surveys aiming to detect Jupiter-sized planets,
telescopes with apertures of $D\sim (200-400)~{\rm cm}$ are optimal.
Exposure times of less then a couple minutes are generally sufficient.

These preliminary calculations provide some guide to the best places
and optimal methods to look for planetary transits.  Observers
searching for transits can use the formalism derived here to precisely
determine which systems to search for transits, and what observing
setup to use.

\bigskip

\section{Summary and Discussion} \label{sec:con}

In this paper we have developed a formalism to predict the efficiency
of searches for transits in stellar systems.  We have taken into
account most relevant parameters that affect the number of transits
that can be observed, and we have described how the total number of expected
transit detections depends on these parameters.  Our primary results are
as follows.  

\begin{description}

\item {\bf (1) $I$-band is optimal.}  For the range of parameters encountered
in most transit surveys, observations in the $I$-band maximize the number of
detected planetary transits.  In general, redder bands are preferred also 
because the effects of limb-darkening are minimized, which aids in the
interpretation of transit candidates and in the elimination of false positives
\citep{mo03}.  However we have not taken into account the variation in 
quantum efficiency of detectors as a function of wavelength.  This is likely a small effect.  
For some detectors, fringing can be a serious problem in the $I$-band.  Thus, in some cases,
somewhat bluer bands (i.e.\ the $R$-band) may be preferable.  Surprisingly, in 
essentially no case do we find that observations in the $K$-band outperform those in the $I$-band.

\item {\bf (2) The $\sn$ depends weakly on primary mass.}  For the $I$-band, and assuming
only Poisson uncertainties, the signal-to-noise ratio
of a planetary transit is an extremely weak function of mass for sources
with flux above that of the sky background.  For sources with flux below
sky, the $\sn$ is a very strong decreasing function of mass.

\item {\bf (3) The number of detections is proportional to the number of stars above sky.}  
As a direct consequence of item (2), if one can find planets around any stars in the
target system, one can detect planets around all stars in the system with fluxes
above sky.  Therefore, the number of planets that are detectable is proportional
to the number of stars with flux above sky.  This is quite distinct from
the usual assumption that the number of detectable planets is proportional to
the number of stars with photometric error less than a given precision, usually
taken to be $\sim 1\%$.  Estimates based on this canonical
criteria will typically be incorrect.

\item {\bf (4) Most planets will be detected around stars with flux near sky.}  
Under the typically valid assumption of a mass function that rises toward lower-mass
stars, item (3) implies that most planets will be detected around stars that 
have fluxes approximately equal to the flux of the sky background.

\item {\bf (5) Planets orbiting stars near sky must be detectable.}
The primary requirement for a successful transit survey is that the planets
orbiting stars with fluxes near the sky background must be detectable. 
This requirement is formulated mathematically in \eq{eqn:regime}.
Provided this requirement is met, the number of detected planets is a rather weak
function of the radius of the planet, index of the bolometric mass-luminosity
relation, age of the system, index of the mass function, seeing,
exposure time, telescope aperture, and the detection threshold.

\item {\bf (6) The richest, closest systems are optimal.}
The number of detected planets has the strongest dependence on the distance modulus, the distance
and extinction to the cluster.  Systems at distances of $d\la 2~{\rm kpc}$ are optimal.
Very nearby ($d \la 1~{\rm kpc}$) systems may have difficulties with saturation of bright stars,
as well fitting within the field-of-view of the detector.

\item {\bf (7) Follow-up of the majority of candidates may be
difficult.}  The majority of planets in typical target systems are likely to be detected around stars 
with apparent magnitudes of $V\ga 17$, making
precision RV follow-up difficult.  This is an important conclusion
which may affect the design of transit surveys.  Difficulties with RV
follow-up are partially ameliorated by the fact that surveys toward
stellar systems are much less prone to the ambiguities with the
interpretation of candidate detections encountered in field surveys,
since the properties of the primaries are better known.
\end{description}

There are a number of ways in which our analysis could be expanded
and refined.  For instance, we do not take into account the
metallicity of the observed systems.  Studies have indicated that
planets are more common around high-metallicity stars
\citep{fischer05}, and as that correlation becomes better
characterized we can add metallicity to the parameters we examine.
It would also be useful to include the effects of observability
on the window function, which is important for the selection of 
optimal targets.  Our analysis also does not account for bad weather.  Further work
could examine what kinds of inclement weather are most damaging for a
transit search and could possibly address the question of
whether it is possible to partially compensate for inclement weather 
by adopting more sophisticated observing strategies. Another
potential refinement would be to account for different forms for the
mass function, rather than relying on the simple Salpeter shape we use
in this paper, such as a broken power-law function with different power-law
indices for high mass and low mass stars.  Lastly, we do not account for stellar 
binarity, which also generally decreases detection probability.

\bigskip

\acknowledgments 
We would like to thank Chris Burke for useful discussions.  We would also like 
to thank the referee for a prompt response and helpful suggestions.  This 
work was supported by a Menzel Fellowship from the 
Harvard College Observatory, and also by the National Aeronautics and
Space Administration under Grant No. NNG04GO70G issued through the Origins
of Solar Systems program.

\appendix

\bigskip

\section{Total S/N Formulation} \label{app:totsn}

In the main text, we derived expressions for the signal-to-noise ratio
of a transiting planet $\sn=(\Delta \chi^2)^{1/2}$ for a single transit. 
The probability
$\psn$ that a planet will have a $\sn$ that exceeds a given threshold,
and all subsequent calculations, were based on this 
single-transit $\sn$ criterion.  
However, planets will generally exhibit
multiple transits, and it is possible, by folding an observed light
curve about the proper period, to improve the total
$\sn$ over that of a single transit by $\sim n^{1/2}$, where $n$ is
the number of observed transits.  In fact, popular 
transit search algorithms operate on phase-folded light curves,
and so trigger based on this total signal-to-noise ratio \citep{kovacs02,aigrain04,ws05}.
It is therefore interesting to rederive our expressions based on 
this total $\sn$ formulation.

The general expression for the number of detected planets $N_{det}$ 
remains the same, but the expression for the 
total detection probability $\ptot(M,P,r)$ needs to be altered,
\begin{equation}
\ptot(M,P,r) = \pt(M,P) \psn^{tot}(M,P,r) \pw(P),
\label{eqn:pall2}
\end{equation}
where $\pt$ and $\pw$ are the transit and window probabilities as before, 
and $\psn^{tot}$ is now
the probability that the {\it total} signal-to-noise ratio
is higher than some threshold value.  

The total signal-to-noise probability, $\psn^{tot}$ 
can be derived in an analogous
way as the one-transit signal-to-noise probability 
(see \S\ref{sec:psn}).  We begin by defining
$d\psn^{tot}/db\equiv \Theta \left[ 
\Delta\chi^2_{tr} -  \Delta\chi_{min}^2 \right]$,
where $\Delta\chi^2_{tr}$ is the difference in $\chi^2$ between a constant flux and transit
fit to the data,
\begin{equation}
\Delta\chi^2_{tr}=N_{tr}^{tot} \left( \frac{\delta}{\sigma} \right)^{2}.
\label{eqn:chitrtot}
\end{equation}
Here $N_{tr}^{tot}$ is the total number of observations taken during 
{\it any} transit, and $\delta$ and  $\sigma$ are as before.

For no aliasing, and periods much shorter
than the length of the observational campaign, the total number of observations during
transit is simply the transit duty cycle $t_{tr}/P$, 
times the total number of observations $N_{tot}$,
\begin{equation}
N_{tr}^{tot}= \frac{t_{tr}}{P}N_{tot}.
\label{eqn:ntrtot}
\end{equation}
In fact, for campaigns of finite durations from single sites, aliasing cannot
be ignored, and there will be a dispersion in the fraction
of points during transit about the naive estimate $t_{tr}/P$.  For
long campaigns lasting more than $\sim 40~{\rm days}$, aliasing effects are
generally not dominant, although they are still significant (see \citealt{gsm05} for 
examples).  They can be accounted for by integrating $d^2\psn/d\phi db$ over the transit
phase $\phi$ as well as impact parameter $b$.  For simplicity,
we will ignore aliasing effects here, and assume \eq{eqn:ntrtot}. 
 
Since $t_{tr}=\sqrt{1-b^2} t_{eq}$, we can write,
\begin{equation}
\Delta \chi^2_{tr}=\Delta \chi^2_{eq}\sqrt{1-b^2},\qquad
\Delta \chi^2_{eq}=N_{tot} \frac{t_{eq}}{P}\left( \frac{\delta}{\sigma} \right)^{2},
\label{eqn:dchitreq}.
\end{equation}
The total signal-to-noise probability is then just the integral over impact
parameter,
\begin{equation}  \label{eqn:psn1b2a}
\psn^{tot}= \int_0^1 \frac{d\psn}{db} db,
\end{equation}
which yields
\begin{equation}  \label{eqn:psn1b2b}
\psn^{tot}= \sqrt{1- \left( \frac{\Delta\chi^2_{min}}{\Delta\chi^2_{eq}} \right)^2}\,,
\end{equation}
if $\Delta\chi^2_{min} \le \Delta \chi^2_{eq}$, and $\psn=0$ otherwise.

We can write $\chi^2_{eq}$ in more explicit terms using the expressions
for $\sigma$, $\delta$, and $t_{eq}$ derived previously.  The additional
 new ingredient is the expression for $N_{tot}$.  If we assume
that the campaign lasts $\nn$ nights, each with a duration of $\tn$,
and that observations are made continuously, then the total number of data points
is 
\begin{equation}
N_{tot} = \frac{\tn}{t_{exp}+t_{read}}\nn.
\label{eqn:ntot}
\end{equation}
Combining this with the expressions we derived in \S\ref{sec:psn},
we arrive at the expression,
$$
\Delta\chi^2_{eq} = 
(1024\pi)^{-1/3} 
\frac{t_{exp}}{t_{read}+t_{exp}}
\left(\frac{r}{R}\right)^4 
\left(\frac{D}{d}\right)^2 
\left(\frac{R^3}{GMP^2}\right)^{1/3} 
\tn\nn L_\lambda 10^{-0.4A_\lambda}
$$
\begin{equation}
\times\left(1+\frac{S_{sky,\lambda}\Omega 4\pi d^2}{L_\lambda 10^{-0.4A_\lambda}}\right)^{-1},
\label{eqn:chi2eqtot}
\end{equation}
which can be compared to the analogous expression for a single transit,
\eq{eqn:beta1b}.  Comparison of \eq{eqn:beta1b}
and \eq{eqn:chi2eqtot} reveals that the ratio of $\chi^2_{eq}$ for the
total signal-to-noise ratio formulation to $\chi^2_{eq}$  for the single-transit formulation is $\propto P^{-1}$.  Thus
the total S/N formulation favors short-period planets more heavily than
the single-transit formulation. 

\bigskip

\section{Effect of Partial Transits} \label{app:PartTrans}

Let us return for the moment to our definition of $N_{tr}$.  This variable 
represents the number of observations of the system during a single
transit.  We stated earlier that $N_{tr} = t_{tr}/( t_{read} + t_{exp} )$.  However, 
that formula is only valid if the entire transit is observed during the 
night; it does not hold if only partial transits are observed, i.e.\ 
if the transit begins before the start of the night
or ends after the end of the night.  In those cases the transit is observed 
for a time less than $\ttr$, the number of observations during transit
is less than $\ntr$,
and therefore the signal-to-noise is less than the naive estimate 
in \S\ref{sec:psn}.  

To account for partial transits, we rewrite the transit duration as,
\begin{equation}
\ttr =  \teq \sqrt{1-b^2} f(\phi),
\label{eqn:ttrnew}
\end{equation}
where $f(\phi)$ is the fraction of the total transit duration that occurs during the 
observation window, as a function of the phase $\phi$ of the transit.  For uniform 
sampling, and $\teq\le \tn$, this is simply,
\begin{equation}
f(\phi)=\left\{
\begin{array}{ll}
\frac{1}{2}+\phi\frac{\tn}{\teq\sqrt{1-b^2}}\qquad &{\rm if}\qquad
0\le \phi \le \frac{1}{2}\frac{\teq\sqrt{1-b^2}}{\tn}\\
1 \qquad &{\rm if} \qquad |\phi-\frac{1}{2}| \ge \frac{1}{2}\frac{\teq\sqrt{1-b^2}}{\tn}-\frac{1}{2}\\
\frac{1}{2}+\frac{\tn}{\teq\sqrt{1-b^2}}(1-\phi)\qquad &{\rm if}\qquad 
\phi \ge 1-\frac{1}{2}\frac{\teq\sqrt{1-b^2}}{\tn} \ge 1
\\
0\qquad &{\rm otherwise}
\end{array}\right., 
\end{equation}
where $\phi=0$ is the beginning of the night and $\phi=1$ 
is the end of the night.  Note also that we have also conservatively assumed
that a transit cannot be detected if it is observed for less than half of its
total duration.

\begin{figure}[t]
\epsscale{1.0}
\plotone{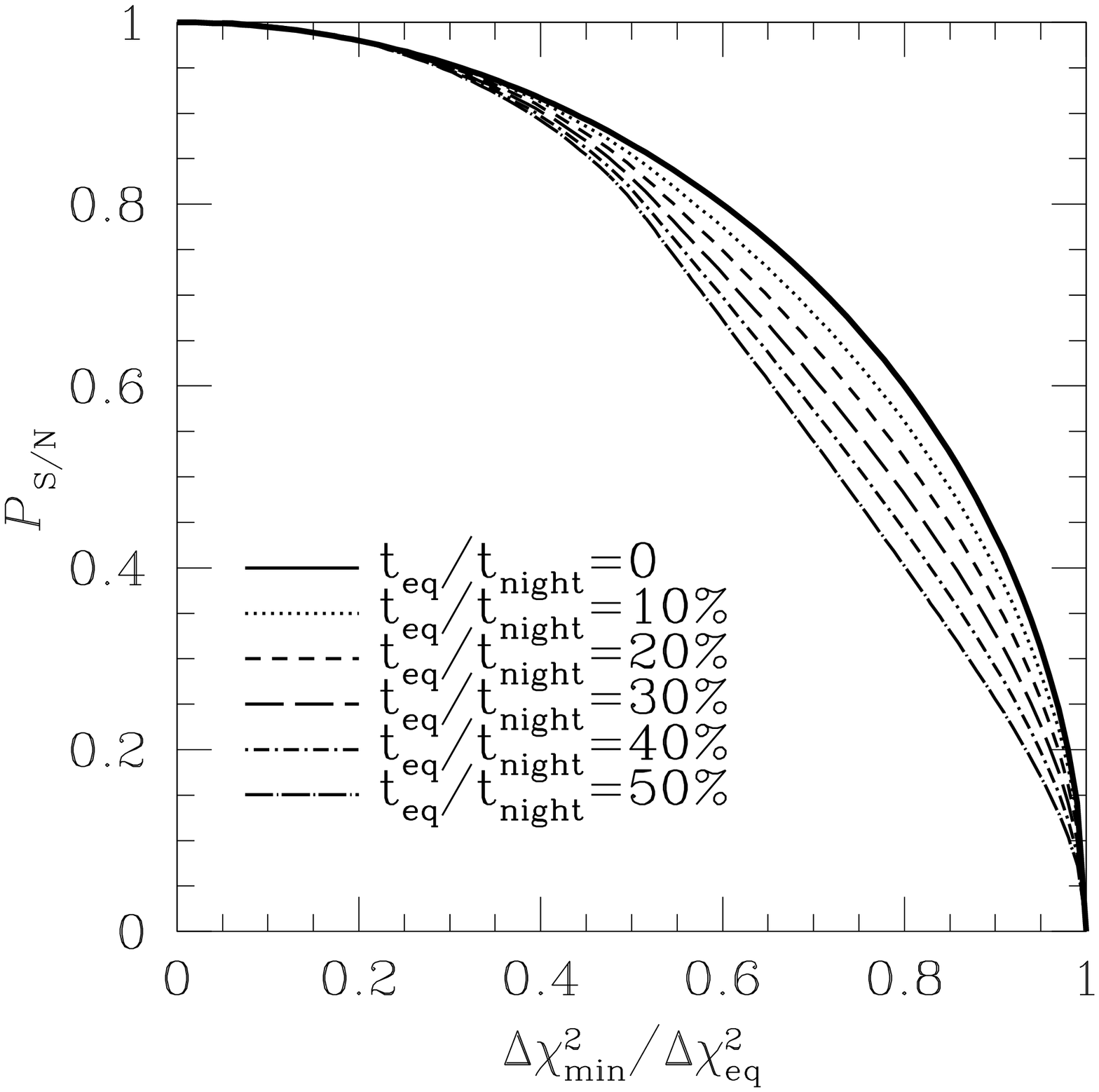}
\caption{The probability $\psn$ that a planet, producing $\Delta\chi^2=\Delta\ceq$ 
for an equatorial transit, will yield a $\Delta\chi^2$ greater than
a given threshold $\cmin$, when integrated over all impact parameters
and phases, for various values of the ratio of the equatorial
transit duration $\teq$ to the duration of the night $\tn$.  The solid
line shows the approximation $\psn=\sqrt{1-(\cmin/\Delta\ceq)^2}$ used in the main test, 
which is valid for $\teq\ll \tn$}
\label{fig:fapp1}
\end{figure}

Following the discussion in \S\ref{sec:psn}, we write
\begin{equation}  \label{eqn:psn2}
\frac{d^2\psn}{db d\phi} = \Theta \left[ \Delta\chi_{eq}^2\sqrt{1
- b^2}f(\phi) - \Delta\chi_{min}^2 \right].
\end{equation}
Proceeding in the same way as in \S \ref{sec:psn}, we integrate
equation (\ref{eqn:psn2}) over $b$ from 0 to $b_{max}$, and $\phi$
from 0 to 1, assuming a uniform distribution for $b$ and $\phi$, solve
for $b_{max}$, i.e.,
\begin{equation} \label{eqn:psneval}
\psn = \int_0^1 db \int_0^1 d\phi \frac{d\psn}{db d\phi}.
\end{equation}
We do not attempt to solve \eq{eqn:psneval} analytically, rather we
evaluate it numerically, noting that $\psn$ depends only on the ratios
$\cmin/\ceq$ and $\teq/\tn$.  Figure \ref{fig:fapp1} shows $\psn$ as a
function of $\cmin/\ceq$ for equatorial transit durations lasting
$10-50\%$ of the night.  We also show the result for the simplified
assumption of $\teq\ll \tn$ that we adopted throughout.  We conclude
that our simple assumption is sufficient for purposes, but note that
it overestimates $\psn$ by as much as $25\%$ for
certain combinations of parameters.

\end{document}